\documentclass[12pt,draftclsnofoot,onecolumn]{IEEEtran}
\usepackage[utf8]{inputenc} 
\usepackage[T1]{fontenc}
\usepackage{url}              
\usepackage{cite}             

\usepackage[cmex10]{amsmath}  
\usepackage{mleftright}       
\mleftright                   

\usepackage{graphicx}         
\usepackage{booktabs}         
                              
\usepackage{bm}
\usepackage{graphicx}
\usepackage{amssymb,amsthm,latexsym,amscd,amsfonts,mathrsfs}
\usepackage{subfigure}
\usepackage{mathtools}
\usepackage{epsfig}
\usepackage{float}
\usepackage{lipsum}
\usepackage{stfloats}
\usepackage{array}
\usepackage{algorithm2e}
\usepackage{algorithmic}
\usepackage{multirow}
\usepackage{fancyh}
\usepackage{upgreek}
\usepackage{dsfont}
\usepackage{psfrag}

\usepackage{multicol, blindtext}
\usepackage{mathtools, cuted}

\usepackage{color}

\usepackage{authblk}
\usepackage{hyperref}
\newtheorem{theorem}{Theorem}
\newtheorem{lemma}{Lemma}

\newtheorem{definition}{Definition}

\newtheorem{corollary}{Corollary}
\usepackage{float}
\usepackage{relsize}
\usepackage{subfigure}
\usepackage{array}
\usepackage{tabularx}
\usepackage{multirow}
\usepackage{tikz}
\newcommand*\circled[1]{\tikz[baseline=(char.base)]{
            \node[shape=circle,draw,inner sep=0.35pt] (char) {#1};}}

\DeclarePairedDelimiter\floor{\lfloor}{\rfloor}
\newcommand\fr[1]{\left\{#1\right\}}
\normalsize

\newtheorem{remark}{Remark}

\DeclarePairedDelimiterX{\Rdivx}[2]{(}{)}{%
  #1\;\delimsize\|\;#2%
}

\begin{document}

\title{Output Statistics of Random Binning: Tsallis Divergence and Its Applications}


\author[$\ast$]{Masoud Kavian}
\author[$\dagger$]{Mohammad M. Mojahedian}
\author[$\dagger$]{Mohammad H. Yassaee}
\author[$\dagger$]{\\Mahtab Mirmohseni,}
\author[$\dagger$]{Mohammad Reza Aref\thanks{This paper was presented in part at ITW 2024.}}
\affil[$\ast$]{\footnotesize Paris Research Center, Huawei Technologies, Paris, France}
\affil[$\dagger$]{\footnotesize Information Systems and Security Lab. (ISSL), Sharif University of Technology, Tehran, Iran}
\affil[ ]{masoud.kavian@huawei.com, m.mojahedian@gmail.com, yassaee@gmail.com, m.mirmohseni@surrey.ac.uk, aref@sharif.edu}

\maketitle

\begin{abstract}
Random binning is a widely used technique in information theory with diverse applications. In this paper, we focus on the output statistics of random binning (OSRB) using the Tsallis divergence $T_\alpha$. We analyze all values of $\alpha \in (0, \infty)\cup\{\infty\}$ and consider three scenarios: (i) the binned sequence is generated i.i.d., (ii) the sequence is randomly chosen from an $\epsilon$-typical set, and (iii) the sequence originates from an $\epsilon$-typical set and is passed through a non-memoryless virtual channel. Our proofs cover both achievability and converse results. To address the unbounded nature of $T_\infty$, we extend the OSRB framework using R\'enyi’s divergence with order infinity, denoted $D_\infty$. As part of our exploration, we analyze a specific form of R\'enyi’s conditional entropy and its properties. Additionally, we demonstrate the application of this framework in deriving achievability results for the wiretap channel, where Tsallis divergence serves as a security measure. The secure rate we obtain through the OSRB analysis matches the secure capacity for $\alpha \in (0, 2]\cup\{{\infty}\}$ and serves as a potential candidate for the secure capacity when $\alpha \in (2, \infty)$.
\end{abstract}

\section{Introduction}
\label{sec:intro}

The concepts of distance between probability measures and correlation metrics are closely intertwined and widely utilized in information theory and machine learning. These metrics play a crucial role in various applications, such as quantifying security levels \cite{mojahedian2019correlation,yu2018renyi,Issa, Kamath,Li,Cuff,Weinberger,bellare2012semantic,dodis2005entropic} and bounding generalization errors \cite{xu2017information,steinke2020reasoning,
esposito2021generalization,rodriguez2021tighter}. Several well-known distance measures are commonly employed, including the total variation distance, KL divergence, R\'enyi divergence, and Tsallis divergence. The total variation distance is a popular metric defined as
\begin{align}
\|p(x)-q(x)\|_{\mathrm{TV}}=\frac 12\sum_x\big|p(x)-q(x)\big|.
\end{align}
KL divergence, another widely used measure, is defined as
\begin{align}
D\big(p(x)\parallel q(x)\big)=\sum_x p(x)\log\left(\frac{p(x)}{q(x)}\right).
\end{align}
R\'enyi divergence generalizes KL divergence and is defined as
\begin{align}
D_\alpha\big(p(x)\parallel q(x)\big)=\frac{1}{\alpha-1}\,\log\left(\sum_x p(x)^\alpha q(x)^{1-\alpha}\right),\label{Renyi:divergenve:def}
\end{align}
where $\alpha$ is a parameter that determines the order of the divergence.

Tsallis divergence is another generalization of relative entropy, given by
\begin{align}
T_{\alpha}\left(p(x)\parallel q(x)\right)=\frac{1}{\alpha-1}\left(\sum_{x}p(x)^{\alpha}q(x)^{1-\alpha}-1\right).\label{eqn:tsallis_divergence}
\end{align}

Correlation measures also play a significant role in various scientific domains. Two well-known measures are Shannon's mutual information and R\'enyi mutual information. Shannon's mutual information is defined as
\begin{align*}
I(X;Y)=\sum_{x,y}p(x,y)\log\frac{p(x,y)}{p(x) p(y)}.
\end{align*}
R\'enyi mutual information of order $\alpha$, proposed by Sibson, is defined as
\begin{align*}
I_{\alpha}(X;Y)&=\frac{\alpha}{\alpha-1}\,\log\left(\sum_y \left[\sum_x p(x)p(y|x)^{\alpha}\right]^{1/\alpha}\right),
\end{align*}
where $\alpha$ belongs to the range $(0,1)\cup(1,\infty)$.

These different metrics exhibit various relationships with each other, with some metrics being stronger or weaker than others. Bounding one metric often leads to bounding another. For instance, Pinsker's inequality states that mutual information is a stronger measure than total variation distance, as shown by the inequality:
\begin{align}
\|p(x,y)-p(x)p(y)\|_\mathrm{TV}\leq\sqrt{\frac{1}{2}I(X;Y)}.
\end{align}
Moreover, R\'enyi mutual information is a non-decreasing function of its order, implying that any upper bound on R\'enyi mutual information with an order greater than $1$ yields a bound on mutual information.

Random binning has long been recognized as a valuable tool in information theory, particularly for establishing achievability results. One essential technique for analyzing random binning is the method introduced in \cite{yassaee2014achievability}. This work investigates the output statistics of random binning (OSRB) and demonstrates that, when the binning rate is below a certain threshold, the bin index variable becomes asymptotically independent from other variables as the block length increases, based on the total variation criterion. In \cite{MoGoAr16}, the asymptotic analysis of OSRB is extended to a stronger measure of mutual information.

We consistently work with random variables where their correlations are significant, ranging from message and ciphertext in security contexts to sample and hypothesis spaces in statistical learning theory. This paper focuses on the Tsallis correlation measure, encompassing various measures such as mutual information. We introduce an OSRB-based tool capable of analyzing scenarios where this measure is applicable. Specifically, by bounding the Tsallis measure, we demonstrate that the binning rate will be constrained by a conditional R\'enyi entropy. Several definitions of conditional R\'enyi entropy are conceivable \cite{hayashi2011exponential,vskoric2011sharp,fehr2014conditional,renyi1961measures,sakai2020asymptotic}. The conditional R\'enyi entropy employed in this paper has already been introduced in \cite{hayashi2011exponential}. But in this paper, this conditional R\'enyi entropy naturally results from OSRB analysis based on Tsallis divergence, which gives it an operational meaning. Also, in this paper, the properties of the proposed conditional R\'enyi entropy are well investigated. 

We present OSRB theorems for three scenarios: one in which the binned sequence is generated as i.i.d., another where it is randomly chosen from an $\epsilon$-typical set, and a third scenario in which the binned sequence comes from an $\epsilon$-typical set and is then passed through a non-memoryless virtual channel. We analyze the second and third cases as deterministic and stochastic encoders, respectively. As an application, we derive a secure rate for the wiretap channel using the Tsallis divergence as the security criterion. By utilizing the OSRB theorem for both deterministic and stochastic cases, we obtain a higher achievable rate. Our findings can be viewed as an extension of \cite{yu2018renyi}, where the secure rate of the wiretap channel is computed using R\'enyi divergence with parameter $\alpha \in (0,2] \cup {\infty}$ as the measure of secrecy.

The remaining sections of this paper are structured as follows. Section \ref{sec:Preliminaries} provides the necessary preliminaries. In Section \ref{sec:cond_renyi_entropy}, we introduce a specific type of conditional R\'enyi entropy and investigate its properties. Theorems for asymptotic OSRB based on the Tsallis measure criterion are presented in Section \ref{sec:renyi_statistics}. Section \ref{sec:applications} explores the applications of Tsallis-based OSRB in analyzing the achievability rate regions of the wiretap channel problem. Finally, Section \ref{sec:conclusion} presents the concluding remarks of the paper.

\section{Preliminaries}
\label{sec:Preliminaries}

This section begins by introducing the relevant notations and providing a definition of random binning. We then identify a counting problem and discuss the necessary properties of Tsallis divergence.

\subsection{Notations and Definitions}
Random variables are denoted by capital letters and their values by lowercase letters. Alphabet sets of random variables are shown in calligraphic font. The $n$-ary Cartesian power of a set $\mathcal{X}$ is denoted as $\mathcal{X}^n$. The notation $p^{U}\!(x)$ represents the uniform distribution over the set $\mathcal{X}$. We use $[\ell]$ to denote the set $\{1,2,\dots, \ell\}$.  We use $\mathds{1}\{\cdot\}$ to denote the indicator function; it is equal to one if the condition inside $\{\cdot\}$ holds; otherwise, it is zero. For a real number $x$, its integer part and fractional part are  denoted by $\floor{x}$ and $\fr{x}$, respectively. All the logarithms in this paper are in base two. 

\begin{definition}[Random binning]
In the process of random binning, each realization of a random variable is randomly assigned to a bin index. Therefore, random binning can be viewed as a random function denoted as $\mathcal{B}:\mathcal{X}\rightarrow[M]$, where each symbol $x\in\mathcal{X}$ is uniformly and independently mapped to a symbol $b\in[M]$. Consider two dependent random variables $X$ and $Z$ with a joint probability mass function $p(x,z)$. By applying random binning, we map the set $\mathcal{X}$ to $[M]$. As a result, the induced random probability mass function on the set $\mathcal{X}\times\mathcal{Z}\times[M]$ can be expressed as
\begin{align}
P(x,z,b) = p_{X,Z}(x,z)\,\mathds{1}\{\mathcal{B}(x)=b\}.
\end{align}
Consequently, we obtain the conditional probability mass function as
\begin{align}
\label{eqn:rand_bin_dist}
P(b\lvert z) = \sum_x p(x\lvert z)\,\mathds{1}\{\mathcal{B}(x)=b\}.
\end{align}
Here, the use of capital letter $P$ indicates that the induced probability mass function on $x,z,b$ is random.
\end{definition}

\begin{definition}
A channel $p(x|z)$ is referred to as a singleton channel if, for any $x_1, x_2, z_1, z_2$ such that $p(z_1)p(z_2)>0$, the following condition holds.
\begin{align}
p(x_1|z_1) = p(x_2|z_2).
\end{align}
On the other hand, if the above condition is not satisfied, the channel is classified as a non-singleton channel.
\end{definition}

\subsection{Counting}
The count of positive integer solutions for the linear equation
\begin{align}
x_1+x_2+\ldots+x_\ell=\alpha,
\end{align}
is given by
\begin{align}
\binom{\alpha-1}{\ell-1}=\frac{(\alpha-1)!}{(\ell-1)!(\alpha-\ell)!}.
\end{align}
It is evident that the following inequality holds.
\begin{align}
\label{coefficient}
\binom{\alpha-1}{\ell-1}\leq\alpha^\alpha.
\end{align}

\subsection{Some useful properties of Tsallis divergence}
By using the inequality $\log x\leq x-1$, we can establish the following inequalities for Tsallis and R\'enyi divergences.
\begin{align}
T_{\alpha}\left(p(x)\parallel q(x)\right)&\geq D_{\alpha}\left(p(x)\parallel q(x)\right)\qquad\alpha\in(1,\infty)\label{Tisalis:D:Compare1}\\
D_{\alpha}\left(p(x)\parallel q(x)\right)&\geq T_{\alpha}\left(p(x)\parallel q(x)\right)\qquad\alpha\in(0,1).\label{Tisalis:D:Compare2}
\end{align}
Moreover, the limit values of the Tsallis divergence are given by
\begin{align}
\lim_{\alpha\to 1}T_{\alpha}\left(p(x)\parallel q(x)\right)&=D\left(p(x)\parallel q(x)\right),\label{Tisalis:properties:1}
\end{align}
and
\begin{align}
\lim_{\alpha\to\infty}T_{\alpha}\left(p(x)\parallel q(x)\right)&=\infty\label{Tisalis:properties:2},
\end{align}
for two different distributions $p(x)$ and $q(x)$.

Due to the unbounded nature of the Tsallis divergence as $\alpha$ approaches infinity, it is less appealing for investigation. Therefore, the paper focuses on analyzing $D_\infty$ instead, defined as
\begin{align}
D_{\infty}\left(p(x)\parallel q(x)\right)=\log\left(\max_{x}\frac{p(x)}{q(x)}\right).\label{Tisalis:properties:3}
\end{align}

\begin{remark}
Considering the definitions of $T_\alpha$ and $D_\alpha$, represented as $x-1$ and $\log(x)$, respectively, when one of them tends to $0$ or $\infty$, the other follows suit.
\end{remark}

\section{Conditional R\'enyi entropy}
\label{sec:cond_renyi_entropy}

In this section, we present a candidate for the conditional R\'enyi entropy and thoroughly examine its properties. This particular entropy emerges naturally when bounding the output of random binning using the Tsallis divergence. Since the properties of this entropy will be utilized in the subsequent theorems, we introduce it earlier. Although the conditional R\'enyi entropy we introduce in this paper has been previously discussed in \cite{hayashi2011exponential} and some of its properties have been investigated in \cite{iwamoto2013revisiting}, we provide a comprehensive overview of all its properties here for the sake of completeness.

\begin{lemma}
\label{decreasinglemma}
Consider the function 
\begin{align}
\label{eqn:cond_renyi_ent}
\tilde{H}_{\alpha}(X|Z)=\frac{1}{1-\alpha}\log\left(\sum_{z}p(z)\sum_{x}p^{\alpha}(x|z)\right).
\end{align}
Then it has the following properties

\begin{enumerate}
\item Let $p(x,z)=p(x)p(z)$, then
\begin{align}
\tilde{H}_{\alpha}(X|Z)=H_\alpha(X),
\end{align}
where $H_\alpha(X)$ is the R\'enyi entropy of order $\alpha$.
\item\label{prop:second} $\tilde{H}_{\alpha}(X|Z)$ is decreasing in $\alpha\in\mathbb{R}$. To be more precise, it is strictly decreasing for non-singleton channel $p(x|z)$.

\item\label{prop:alpha1}$\lim_{\alpha\to 1}\tilde{H}_{\alpha}(X|Z)=H(X|Z)$.

\item Data processing inequality for $\alpha\in(1,\infty)$,
\begin{align}
\tilde{H}_{\alpha}(X|Y)\leq \tilde{H}_{\alpha}(X|Z),
\end{align}
such that $X-Y-Z$ forms a Markov chain.
\item Suppose $\alpha\in(1,\infty)$ and $\lambda\in[0,1]$, then $\tilde{H}_{\alpha}(X|Z)$ for fixed $p(x|z)$ is a convex function of $p(z)$. More precisely, we have
\begin{align}
&\frac{1}{1-\alpha}\log\left(\sum_{z}\Big(\lambda p_{\lambda}(z)+\bar{\lambda}p_{\bar{\lambda}}(z)\Big)\cdot\sum_{x}p^{\alpha}(x|z)\right)\nonumber\\
&\qquad\leq\frac{\lambda}{1-\alpha}\log\left(\sum_{z}p_{\lambda}(z)\sum_{x}p^{\alpha}(x|z)\right)+\frac{\bar{\lambda}}{1-\alpha}\log\left(\sum_{z}p_{\bar{\lambda}}(z)\sum_{x}p^{\alpha}(x|z)\right),
\end{align}
where $\bar{\lambda}=1-\lambda$ and $p_{\lambda}(z)$ and $p_{\bar{\lambda}}(z)$ are two different probability mass functions over the alphabet $\mathcal{Z}$.

\item In the limit as $\alpha\rightarrow\pm\infty$,\label{prop:infty} 
\begin{align}
\lim_{\alpha\to\infty}\tilde{H}_{\alpha}(X|Z)&=\log\left(\frac{1}{\max_{x,z}p(x|z)}\right)\\
\lim_{\alpha\to-\infty}\tilde{H}_{\alpha}(X|Z)&=\log\left(\frac{1}{\min_{x,z}p(x|z)}\right).
\end{align}
\item Let $(X^n,Z^n)$ be i.i.d. random variables distributed according to $p(x,z)$, i.e.,
\begin{align}
p(x^n,z^n)=\prod_{i=1}^n p(x_i,z_i).
\end{align}
Then,
\begin{align}
\tilde{H}_{\alpha}(X^n|Z^n)=\sum_{i=1}^n\tilde{H}_{\alpha}(X_i|Z_i)=n\tilde{H}_{\alpha}(X|Z).
\end{align}

\item $\tilde{H}_{\alpha}(X|Z)$ is non-negative.

\item \label{prop:equality_s}For singleton channel $p(x|z)$ and any $\alpha_1\neq\alpha_2\in\mathbb{R}$, we have
\begin{align}
\tilde{H}_{\alpha_1}(X\lvert Z)=\tilde{H}_{\alpha_2}(X\lvert Z).
\end{align}
\end{enumerate}

\end{lemma}
The proofs of the properties are given in Appendix \ref{subsec:app_prop}.
\begin{lemma}[Harris' inequality]
Let $f, g: \mathbb{R} \to \mathbb{R}$ be non-decreasing functions, and let $X$ be a real-valued random variable taking values in $\mathbb{R}$. Then  
\begin{align}
\mathbb{E}\left[f(X) g(X)\right]\geq \mathbb{E}\left[f(X)\right]\,\mathbb{E}\left[g(X)\right],
\end{align}
which by induction and paying attention to the fact that the multiplication of non-decreasing and non-negative functions is non-decreasing, we have
\begin{align}
\mathbb{E}\left[\prod_{i=1}^{\ell}f_i(X)\right]\geq\prod_{i=1}^{\ell}\mathbb{E}\left[f_i(X)\right]
\end{align} 
when $f_i(X)$ is non-decreasing and non-negative.
\end{lemma}
\section{Statistics of random binning}
\label{sec:renyi_statistics}

In the subsequent analysis, we examine the output statistics of random binning based on the Tsallis divergence criterion. We consider two separate cases, namely $\alpha\in(0,1)$ and $\alpha\in(1,\infty)$, and present two theorems that encompass both achievability and converse proofs. Additionally, we provide a theorem specifically for the case of $\alpha=\infty$, where we utilize $D_\infty$ as a measure.

\subsection{Statistics of random binning applied to i.i.d. random variables}
In this subsection, we assume that $X^n\sim\prod_{i=1}^{n}p(x_i)$.
\begin{theorem}
\label{RenyiOSRB}
	Consider the scenario where $\alpha\in(1,\infty)$ and $\mathcal{B}:\mathcal{X}^n\rightarrow[2^{nR}]$ represents the set of all random mappings with a rate of $R$, satisfying
	\begin{align}
		R<\tilde{H}_{\alpha}(X|Z).
	\end{align}
	As $n$ approaches infinity, we observe that
	\begin{align}
		\mathbb{E}_{\mathcal{B}}&\left[T_\alpha\Big(P(b,z^n)\parallel p^{U}\!(b)\,p(z^n)\Big)\right]\to 0.
	\end{align}
	Furthermore, if $R>\tilde{H}_{\alpha}(X|Z)$, then we have
	\begin{align}
		\mathbb{E}_{\mathcal{B}}&\left[T_\alpha\Big(P(b,z^n)\parallel p^{U}\!(b)\,p(z^n)\Big)\right]\to \infty.
	\end{align}
\end{theorem}

\begin{proof}
We begin by proving the theorem for $\alpha\in\mathbb{N}$, and then we proceed to generalize it. It is important to note that the proof initially focuses on the single-shot scenario and later extends to the asymptotic regime. By considering the definition of Tsallis divergence in equation \eqref{eqn:tsallis_divergence}, we obtain
\begin{align}
\mathbb{E}_{\mathcal{B}}\left[T_\alpha\Big(P(b,z)\parallel p^{U}\!(b)\,p(z)\Big)\right]&=\frac{1}{\alpha-1}\left(\mathbb{E}_{\mathcal{B}}\left[M^{\alpha-1}\sum_{b,z}P^\alpha(b|z)p(z)\right]-1\right)\\
&=\frac{1}{\alpha-1}\left(\sum_{z}p(z)\left(M^\alpha\mathbb{E}_{\mathcal{B}}\Big[P^\alpha(b=1|z)\Big]\right)-1\right)\label{eq:expec},
\end{align}
where \eqref{eq:expec} is due to the symmetry and linearity of expectation.

Using the multinomial expansion, we now compute the expectation by substituting \eqref{eqn:rand_bin_dist} in \eqref{eq:expec}.
\begin{align}
&M^\alpha\mathbb{E}_{\mathcal{B}}\Big[p(b=1|z)^{\alpha}\Big]\\
&\quad=M^\alpha\mathbb{E}_{\mathcal{B}}\left[\bigg(\sum_{x}p(x\lvert z)\mathds{1}\{\mathcal{B}(x)=1\}\bigg)^\alpha\right]\label{expand1}\\
&\quad=M^\alpha\mathbb{E}_{\mathcal{B}}\left[\sum_{x_1,\ldots,x_\alpha}\prod_{i=1}^{\alpha}p(x_i\lvert z) \mathds{1}\{\mathcal{B}(x_i)=1\}\right]\\
&\quad=M^\alpha\mathbb{E}_{\mathcal{B}}\Bigg[\sum_{x_1}p^{\alpha}(x_1\lvert z)\mathds{1}\{\mathcal{B}(x_1)=1\}\nonumber\\
&\quad\qquad\qquad+\Bigg(\sum_{x_1\neq x_2}p^{\alpha-1}(x_1\lvert z)p(x_2\lvert\nonumber z)\cdot\prod_{i=1}^{2}\mathds{1}\{\mathcal{B}(x_i)=1\}\nonumber\\
&\qquad\qquad\qquad +\sum_{x_1\neq x_2}p^{\alpha-2}(x_1\lvert z)p^2(x_2\lvert z)\cdot\prod_{i=1}^{2}\mathds{1}\{\mathcal{B}(x_i)=1\}+\ldots\Bigg)\nonumber\\
&\quad\qquad\qquad+\Bigg(\sum_{x_1\neq x_2\neq x_3}p^{\alpha-2}(x_1\lvert z)p(x_2\lvert z)p(x_3\lvert z)\cdot\prod_{i=1}^{3}\mathds{1}\{\mathcal{B}(x_i)=1\}\nonumber\\
&\qquad\qquad\qquad+\sum_{x_1\neq x_2\neq x_3}\!\!\!p^{\alpha-3}(x_1\lvert z)p^2(x_2\lvert z)p(x_3\lvert z)\cdot\prod_{i=1}^{3}\mathds{1}\{\mathcal{B}(x_i)=1\}\Bigg)\nonumber\\
&\quad\qquad\qquad+\qquad\qquad\qquad\qquad\qquad\cdots\nonumber\\
&\quad\qquad\qquad+\sum_{x_1\neq\cdots\neq x_\alpha}\prod_{i=1}^{\alpha}p(x_i\lvert z) \mathds{1}\{\mathcal{B}(x_i)=1\}\Bigg]\label{eqn:expand1}\\
&\quad=M^{\alpha-1}\sum_{x_1}p^{\alpha}(x_1\lvert z)\nonumber\\
&\qquad+M^{\alpha-2}\Big[\sum_{x_1\neq x_2}p^{\alpha-1}(x\lvert z)p(x_2\lvert z)+\sum_{x_1\neq x_2}p^{\alpha-2}(x_1\lvert z)p^2(x_2\lvert z^n)+\cdots\Big]\nonumber\\
&\qquad +M^{\alpha-3}\Big[\sum_{x_1\neq x_2\neq x_3}p^{\alpha-2}(x\lvert z)p(x_2\lvert z)p(x_3\lvert z)\nonumber\\
&\qquad\qquad\qquad+\sum_{x_1\neq x_2\neq x_3}\!\!\!p^{\alpha-3}(x\lvert z)p^2(x_2\lvert z)p(x_3\lvert z)+\cdots\Big]\nonumber\\
&\qquad+\cdots+\sum_{x_1\neq x_2\neq\cdots\neq x_\alpha}\prod_{i=1}^{\alpha}p(x_i\lvert z)\label{eqn:bin_indep}\\
&\quad\leq M^{\alpha-1}\sum_{x}p^{\alpha}(x\lvert z)\label{simpleintiger3}\\
&\qquad+\Bigg[M^{\alpha-2}\Big(\sum_{x}p^{\alpha-1}(x\lvert z)\Big)+M^{\alpha-2}\Big(\sum_{x} p^{\alpha-2}(x\lvert z)\Big)\Big(\sum_{x}p^2(x\lvert z)\Big)+\cdots\Bigg]\nonumber\\
&\qquad+\Bigg[M^{\alpha-3}\Big(\sum_{x}p^{\alpha-2}(x\lvert z)\Big)+M^{\alpha-3}\Big(\sum_{x}p^{\alpha-3}(x\lvert z)\Big)\Big(\sum_{x}p^2(x\lvert z)\Big)+\cdots\Bigg]\nonumber\\
&\qquad+\cdots+1,\label{eq:bound_Ren}
\end{align}
where \eqref{eqn:expand1} comes from the fact that the set $\mathcal{X}^\alpha$ can be partitioned as $\cup_{\ell=1}^\alpha\mathcal{A}_\ell$ where $\mathcal{A}_\ell$ is the set of $\alpha$-tuples in the following form 
\begin{align}\label{participating:nautral:numbersLproof}
\mathcal{A}_\ell=\Big\{(\underbrace{x_1,\ldots,x_1}_{\alpha_1},\underbrace{x_2,\ldots,x_2}_{\alpha_2},\ldots,\underbrace{x_\ell,\ldots,x_\ell}_{\alpha_\ell}):&x_j\in\mathcal{X},j\in[\ell],x_i\neq x_j, i\neq j\Big\},
\end{align}
with $\alpha_1+\ldots+\alpha_\ell=\alpha$. In other words, $\mathcal{A}_\ell$ means that we divide the set $\mathcal{X}^\alpha$ into $\ell$ parts and the data inside each part are equal, while they are not equal to the data of other parts.

Counting how the power $\alpha$ is divided into $\ell$ parts yields an upper bound of \eqref{coefficient} for the cardinality of the set $\mathcal{A}_\ell$. The equation \eqref{eqn:bin_indep} is a consequence of the fact that random binning maps independently and uniformly, or equivalently
\begin{align}
\label{eqn:rand_bin_exp}
\mathbb{E}_\mathcal{B}\left[\prod_{i=1}^{k} \mathds{1}\{\mathcal{B}(x_i)=1\}\right] =\frac{1}{M^k}.
\end{align}
\eqref{simpleintiger3} arises from
\begin{align}
\sum_{x_1\neq x_2\neq\cdots\neq x_\ell}\prod_{i=1}^{\ell}p^{\alpha_i}(x_i\lvert z)\leq \prod_{i=1}^{\ell}\sum_{x}p^{\alpha_i}(x\lvert z).
\end{align}
In addition, we benefited from
\begin{align}\label{upper:muliple:prob}
\sum_{x_1\neq\cdots\neq x_\alpha}\prod_{i=1}^{\alpha}p(x_i\lvert z)\leq \sum_{x_1,\ldots, x_\alpha}\prod_{i=1}^{\alpha}p(x_i\lvert z)=1.
\end{align}
The terms in \eqref{eq:bound_Ren}, by Harris' inequality, can be further bounded as
\begin{align}
M^{\alpha-\ell}\cdot\prod_{i=1}^{\ell}\left(\sum_{x} p^{\alpha_i}(x|z)\right)&=M^{\alpha-\ell}\cdot\prod_{i=1}^{\ell}\mathbb{E}_{p_{X\lvert Z=z}}\Bigg[p^{\alpha_i-1}(X\lvert z)\Bigg]\\
&\leq M^{\alpha-\ell}\cdot\mathbb{E}_{p_{X\lvert Z=z}}\Bigg[\prod_{i=1}^{\ell}p^{\alpha_i-1}(X\lvert z)\Bigg]\\
&=M^{\alpha-\ell}\sum_{x} p^{\alpha-\ell+1}(x|z),
\end{align}
where $\sum_{i=1}^\ell \alpha_i=\alpha$. By substituting these upper bounds into \eqref{eq:expec}, we obtain the following upper-bound terms for Tsallis divergence.
\begin{align}
\label{eqn:upb_form}
M^{\alpha-\ell}\sum_z p(z)\sum_{x}p^{\alpha-\ell+1}(x\lvert z),\qquad \ell=1,\ldots,\alpha-1.
\end{align}
Let $(X^n,Z^n)$ be i.i.d. random variables distributed according to $p(x,z)$. Suppose that we randomly (and uniformly) bin the set $\mathcal{X}^n$ into $M=2^{nR}$ bins. Then, with the increase of $n$, each term forming the upper bound in \eqref{eqn:upb_form} tends to zero if the rate applies in the following inequality.
\begin{align}
R<\tilde{H}_{\alpha-\ell+1}(X\lvert Z),\qquad \ell=1,\ldots,\alpha-1,
\end{align}
or equivalently
\begin{align}
R&<\min_\ell \tilde{H}_{\alpha-\ell+1}(X\lvert Z)\overset{(a)}{=}\tilde{H}_{\alpha}(X\lvert Z),
\end{align}
which $(a)$ results from the non-increasing property of $\tilde{H}_{\alpha}(X\lvert Z)$.

It should be noted that for the singleton case, according to property \ref{prop:equality_s}, the value of $\tilde{H}_{\alpha}(X|Z)$ is constant.

Therefore, with the increase of $n$, the upper bound \eqref{eq:bound_Ren} tends to $1$ and subsequently \eqref{eq:expec} tends to zero. By letting $n$ go to infinity, the proof of the theorem for $\alpha>1\in\mathbb{N}$ is complete.

The generalization of the achievability proof to real values of $\alpha$ and the converse part of the proof can be found in Appendix \ref{sec:app_real}.
\end{proof}

\begin{remark}
Inequality \eqref{Tisalis:D:Compare1} shows that considering $D_\alpha$ instead of $T_\alpha$ in the case of $\alpha\in(1,\infty)$, the result of Theorem \ref{RenyiOSRB} will be valid.
\end{remark}

\begin{remark}
In case of $\alpha=1$, Theorem \ref{RenyiOSRB} shows that if $R<H(X\lvert Z)$ holds, then
\begin{align}
\label{eq:MI_OSRB}
\mathbb{E}_\mathcal{B}\left[I(Z;B)\right]\to 0.
\end{align}
This case can also be deduced from the results of \cite{yassaee2014achievability,MoGoAr16}. More specifically, it has been proven in \cite{yassaee2014achievability} that $R<H(X\lvert Z)$ gives results
\begin{align}
\label{eq:TV_OSRB}
\mathbb{E}_\mathcal{B}\|p(z,b)-p(z)p^{U}(b)\|_{\mathrm{TV}}\to 0.
\end{align}
Furthermore, in \cite{MoGoAr16}, it is shown that \eqref{eq:TV_OSRB} approaches zero exponentially fast, which subsequently results in \eqref{eq:MI_OSRB}.
\end{remark}

The following theorem holds for $\alpha\in(0,1)$.
\begin{theorem}
\label{Theorem:OSRB:alpha:(0,1)}
Let $\alpha\in(0,1)$. Then 
\begin{align}
\mathbb{E}_{\mathcal{B}}&\left[T_\alpha\Big(P(b,z^n)\parallel p^{U}\!(b)\,p(z^n)\Big)\right]\to 0 
\end{align}
holds if and only if the binning rate satisfies $R<{H}(X\lvert Z)$.
\end{theorem}

The proof is provided in Appendix \ref{Proof:OSRB:alpha:(0,1)}.

\begin{theorem}
\label{inftyrandombinning}
If the binning rate satisfies
\begin{align}
R < \tilde{H}_{\infty}(X\lvert Z),
\end{align}
then,
\vspace{-2mm}
\begin{align}
\mathbb{E}_{\mathcal{B}}\left[D_\infty\Big(P(b,z^n)\parallel p^{U}\!(b)\,p(z^n)\Big)\right]\to 0
\end{align}
which $p(z^n)=\prod_{i=1}^{n}p(z_i)$ with $p(z)=\sum_{x}p(x)p(z\lvert x)$.
\end{theorem}

\begin{proof}
For any $0<\beta\leq \dfrac{1}{M\max_{x,z} p(x|z)}$, we have
\begin{align}
&\mathbb{E}_{\mathcal{B}}\left[D_\infty\Big(P(b,z)\parallel p^{U}\!(b)\,p(z)\Big)\right]\\
&=\mathbb{E}_{\mathcal{B}}\left[\log\left(\max_{b,z}MP(b\lvert z)\right)\right]\\
&\leq\log\left(\mathbb{E}_{\mathcal{B}}\left[\max_{b,z}MP(b\lvert z)\right]\right)\label{maxone}\\
&\leq\log\left(\frac{1}{\beta}\log\left(\sum_{b,z}\mathbb{E}_{\mathcal{B}}\Big[\exp\left(\beta MP(b\lvert z)\right)\Big]\right)\right)\label{maxtwo1}\\
&\leq\log\left(\frac{1}{\beta}\log\left(M\lvert\mathcal{Z}\rvert\max_{z}\prod_{x}\left(\frac{e^{M\beta p(x\lvert z)}}{M}+1-\frac{1}{M}\right)\right)\right)\label{maxtwo}\\
&\leq\log\left(\frac{1}{\beta}\log\Bigg(M\rvert\mathcal{Z}\lvert\exp\left(\max_{z}\sum_{x}\frac{e^{M\beta p(x\lvert z)}-1}{M}\right)\right)\label{maxthree}\\
&\leq\log\left(\frac{1}{\beta}\log M\lvert\mathcal{Z}\rvert+\frac{1}{\beta}\max_{z}\Bigg[\beta+\beta^2 M\sum_x p^2(x|z)\Bigg]\right)\label{maxfour}\\
&=\log\left(1+\frac{1}{\beta}\log M\lvert\mathcal{Z}\rvert+\max_{z}\beta M\exp(-H_2(X|Z=z))\right)\\
&\leq \frac{1}{\beta}\log M\lvert\mathcal{Z}\rvert+M\beta \exp(-\min_{z}H_2(X|Z=z))\label{eqn:y-1}\\
&\leq\frac{1}{\beta}\log M\lvert\mathcal{Z}\rvert+M\beta \exp(-\tilde{H}_{\infty}(X|Z)).\label{eqn:y-3}
\end{align}
\eqref{maxone} comes from Jensen's inequality for $f(x)=\log x$. \eqref{maxtwo1} is the consequence of the following inequalities for $\beta\in\mathbb{R}^{+}$.
\begin{align}
\label{eq:max_sum}
	\mathbb{E}\left[\max_{k=1,\ldots,K}Z_k\right]
	&\leq\frac{1}{\beta}\log\left(\sum_{k=1}^{K}\mathbb{E}\Big[\exp(\beta Z_k)\Big]\right).
\end{align}
Replace \eqref{eqn:rand_bin_dist} in \eqref{maxtwo1}, and since we have an independent and uniform mapping, we get \eqref{maxtwo}. \eqref{maxthree} arises from $1+x\leq\exp(x)$, \eqref{maxfour} employs $\exp(x)\leq 1+x+x^2,\, 0\leq x\leq 1$. \eqref{eqn:y-1} results from $\log(1+x)\leq x$. Finally, \eqref{eqn:y-3} follows from
\begin{align}
-\min_{z}&\,H_2(X|Z=z)\nonumber\\
&=\max_z\log\left(\sum_x p^2(x|z)\right)\\
&=\log\left(\max_z\sum_x p^2(x|z)\right)\\
&\leq \log\left(\max_z\sum_x p(x|z)\max_x p(x|z)\right)\\
&=\log\left(\max_{x,z} p(x|z)\right)=-\tilde{H}_\infty(X|Z).
\end{align}
Taking the minimum of \eqref{eqn:y-3} over $0<\beta\leq\dfrac{1}{M\max_{x,z} p(x|z)}$, we obtain
\begin{align}
	\mathbb{E}_{\mathcal{B}}&\left[D_\infty\Big(p(b,z)\parallel p^{U}\!(b)\,p(z)\Big)\right]\nonumber\\
	&\le 2\sqrt{M\log\left(M|\mathcal{Z}|\right)\exp\left({-\tilde{H}_\infty(X|Z)}\right)}\label{eqn:y00}
\end{align}
provided that
\begin{equation}
M\log\left(M|\mathcal{Z}|\right)\exp(-\tilde{H}_\infty(X|Z))<1.\label{eqn:y1}
\end{equation}
Turning to the asymptotic regime, we  consider $(X^n,Z^n)$ in place of $(X,Z)$ in the one-shot setting and we set $M=2^{nR}$. Then if $R<\tilde{H}_\infty(X|Z)$, the condition \eqref{eqn:y1} is satisfied and
\begin{align}
\mathbb{E}_{\mathcal{B}}&\left[D_\infty\Big(P(b,z^n)\parallel p^{U}\!(b)\,p(z^n)\Big)\right]\nonumber\\
&\le 2\sqrt{n\left(R+\log|\mathcal{Z}|\right)\exp\left({R-\tilde{H}_\infty(X|Z)}\right)},\label{eqn:y2}
\end{align}tending exponentially fast to zero.
\end{proof}

\subsection{Statistics of random binning for specific-type sequences}

Up until now, we assumed that $X^n\sim\prod_{i=1}^{n}p(x_i)$. To align our results with those in \cite{yu2018renyi}, we will now consider the distribution of $X^n$ over the $\epsilon$-typical set,
\begin{align}
\mathcal{T}_{\epsilon}^n\left(p_X\right)=\left\{x^n:\Big\lvert\frac{1}{n}\sum_{i=1}^{n}\mathds{1}\{x_i=x\}-p(x)\Big\rvert<\epsilon\right\},
\end{align}
as
\begin{align}
\tilde{p}(x^n)=\frac{\prod_{i=1}^{n}p(x_i)}{\sum_{x^n\in\mathcal{T}_{\epsilon}^n\left(p_X\right)}\prod_{i=1}^{n}p(x_i)}\label{typical:distribution:defenition:1}.
 \end{align}

\begin{theorem}
\label{OSRB:type:achievability}
Let $\alpha \in (1, \infty)$. Assuming that $X^n \sim \tilde{p}(x^n)$ over $\mathcal{T}_{\epsilon}^n\left(p_X\right)$ and $p_{Z^n\lvert X^n}=\prod_{i=1}^{n}p_{Z_i\lvert X_i}$, if the binning rate satisfies 
\begin{align}
R<H(X)-\sum_{x}p(x)D_{\alpha}\Big(p\left(z\lvert x\right)\parallel p\left(z\right)\Big),
\end{align}
then as $n$ goes to infinity, we have
\begin{align}
\mathbb{E}_{\mathcal{B}}&\left[T_\alpha\Big(P(b,z^n)\parallel p^{U}\!(b)\,q(z^n)\Big)\right]\to 0.
\end{align}
Here, $q(z^n)=\prod_{i=1}^{n}p(z_i)$ with $p(z)=\sum_{x}p(x)p(z\lvert x)$, and $H(X)=-\sum_{x}p(x)\log p(x)$.
\end{theorem}

\begin{proof}
Here, we provide the proof only for $\alpha \in \mathbb{N}$, and its generalization to a real $\alpha$ can be straightforwardly obtained using similar steps to the i.i.d. case. Most of the steps in this part are akin to the proof of the i.i.d. case, except we begin this proof in the asymptotic regime. Considering the definition of Tsallis divergence in \eqref{eqn:tsallis_divergence}, we have
\begin{align}
\mathbb{E}_{\mathcal{B}}\left[T_{\alpha}\left(P(b,z^n)\parallel p^{U}(b)q(z^n)\right)\right]&=\frac{1}{\alpha-1}\left(\mathbb{E}_{\mathcal{B}}\left[M^{\alpha-1}\sum_{b,z^n}q(z^n)\left(\frac{P(b,z^n)}{q(z^n)}\right)^\alpha\right]-1\right)\label{type:osrb:tsallis:simplification:1}\\
&=\frac{1}{\alpha-1}\left(\sum_{z^n}q(z^n)M^\alpha\mathbb{E}_{\mathcal{B}}\left(\frac{P(1,z^n)}{q(z^n)}\right)^\alpha-1\right)\label{type:osrb:tsallis:simplification:2},
\end{align} 
where \eqref{type:osrb:tsallis:simplification:2} comes from symmetry and linearity of expectation.

By substituting \eqref{eqn:rand_bin_dist} into \eqref{type:osrb:tsallis:simplification:2}, the multinomial expansion suggests that
\begin{align}
	&M^\alpha\mathbb{E}_{\mathcal{B}}\left(\frac{P(1,z^n)}{q(z^n)}\right)^\alpha\label{type:osrb:tsallis:simplification:3}\\
	&\quad=M^\alpha\mathbb{E}_{\mathcal{B}}\left[\bigg(\frac{\sum_{x^n\in\mathcal{T}_{\epsilon}^n(p_X)}p(x^n, z^n)\mathds{1}\{\mathcal{B}(x^n)=1\}}{q(z^n)}\bigg)^\alpha\right]\label{type:osrb:tsallis:simplification:4}\\
	&\quad=M^\alpha\mathbb{E}_{\mathcal{B}}\left[\left(\frac{\sum_{x^n_1,\ldots,x^n_\alpha}\prod_{i=1}^{\alpha}p(x^n_i,z^n) \mathds{1}\{\mathcal{B}(x^n_i)=1\}}{q^\alpha(z^n)}\right)\right]\label{type:osrb:tsallis:simplification:5}\\
	&\quad=M^\alpha\mathbb{E}_{\mathcal{B}}\Bigg[\sum_{x^n_1\in\mathcal{T}_{\epsilon}^n(p_X)}\frac{p^{\alpha}(x^n_1,z^n)}{q^\alpha(z^n)}\mathds{1}\{\mathcal{B}(x^n_1)=1\}\nonumber\\
	&\quad\qquad\qquad+\Bigg(\sum_{x^n_1\neq x^n_2}\frac{p^{\alpha-1}(x^n_1, z^n)p(x^n_2,\nonumber z^n)\cdot\prod_{i=1}^{2}\mathds{1}\{\mathcal{B}(x^n_i)=1\}}{q^\alpha(z^n)}\nonumber\\
	&\qquad\qquad\qquad +\sum_{x^n_1\neq x^n_2}\frac{p^{\alpha-2}(x^n_1,z^n)p^2(x^n_2,z^n)\cdot\prod_{i=1}^{2}\mathds{1}\{\mathcal{B}(x^n_i)=1\}}{q^\alpha(z^n)}+\ldots\Bigg)\nonumber\\
	&\quad\qquad\qquad+\Bigg(\sum_{x^n_1\neq x^n_2\neq x^n_3}\frac{p^{\alpha-2}(x^n_1, z^n)p(x^n_2,z^n)p(x^n_3,z^n)\cdot\prod_{i=1}^{3}\mathds{1}\{\mathcal{B}(x^n_i)=1\}}{q^\alpha(z^n)}\nonumber\\
	&\qquad\qquad\qquad+\frac{\sum_{x^n_1\neq x^n_2\neq x^n_3}p^{\alpha-3}(x^n_1,z^n)p^2(x^n_2,z^n)p(x^n_3,z^n)\cdot\prod_{i=1}^{3}\mathds{1}\{\mathcal{B}(x^n_i)=1\}}{q^\alpha(z^n)}\Bigg)\nonumber\\
	&\quad\qquad\qquad+\qquad\qquad\qquad\qquad\qquad\cdots\nonumber\\
	&\quad\qquad\qquad+\frac{\sum_{x^n_1\neq\cdots\neq x^n_\alpha}\prod_{i=1}^{\alpha}p(x^n_i,z^n) \mathds{1}\{\mathcal{B}(x^n_i)=1\}}{q^\alpha(z^n)}\Bigg]\label{type:osrb:tsallis:simplification:6}\\
	&\quad=M^{\alpha-1}\sum_{x^n_1\in\mathcal{T}_{\epsilon}^n(p_X)}\frac{p^{\alpha}(x^n_1,z^n)}{q^\alpha(z^n)}\nonumber\\
	&\qquad+M^{\alpha-2}\Big[\sum_{x^n_1\neq x^n_2}\frac{p^{\alpha-1}(x^n_1,z^n)p(x^n_2,z^n)}{q^\alpha(z^n)}+\sum_{x^n_1\neq x^n_2}\frac{p^{\alpha-2}(x^n_1,z^n)p^2(x^n_2,z^n)}{q^\alpha(z^n)}+\cdots\Big]\nonumber\\
	&\qquad +M^{\alpha-3}\Big[\sum_{x^n_1\neq x^n_2\neq x^n_3}\frac{p^{\alpha-2}(x^n,z^n)p(x^n_2,z^n)p(x^n_3,z^n)}{q^\alpha(z^n)}\nonumber\\
	&\qquad\qquad\qquad+\sum_{x^n_1\neq x^n_2\neq x^n_3}\frac{\!\!\!p^{\alpha-3}(x^n,z^n)p^2(x^n_2,z^n)p(x^n_3,z^n)}{q^\alpha(z^n)}+\cdots\Big]\nonumber\\
	&\qquad+\cdots+\sum_{x^n_1\neq x^n_2\neq\cdots\neq x^n_\alpha}\frac{\prod_{i=1}^{\alpha}p(x^n_i, z^n)}{q^\alpha(z^n)}\label{type:osrb:tsallis:simplification:7}\\
	&\quad\leq M^{\alpha-1}\sum_{x^n\in\mathcal{T}_{\epsilon}^n(p_X)}\frac{p^{\alpha}(x^n,z^n)}{q^\alpha(z^n)}\nonumber\\
	&\qquad+\Bigg[(1+2\delta_n)M^{\alpha-2}\Big(\sum_{x^n\in\mathcal{T}_{\epsilon}^n(p_X)}\frac{p^{\alpha-1}(x^n,z^n)}{q^{\alpha-1}(z^n)}\Big)\nonumber\\
&\qquad\qquad\qquad +M^{\alpha-2}\Big(\sum_{x^n\in\mathcal{T}_{\epsilon}^n(p_X)}\frac{p^{\alpha-2}(x^n,z^n)}{q^{\alpha-2}(z^n)} \Big)\Big(\sum_{x^n\in\mathcal{T}_{\epsilon}^n(p_X)}\frac{p^2(x^n,z^n)}{q^2(z^n)}\Big)+\cdots\Bigg]\nonumber\\
&\qquad+\Bigg[(1+2\delta_n)^2M^{\alpha-3}\Big(\sum_{x^n\in\mathcal{T}_{\epsilon}^n(p_X)}\frac{p^{\alpha-2}(x^n,z^n)}{q^{\alpha-2}(z^n)}\Big)\nonumber\\
&\qquad\qquad\qquad+(1+2\delta_n)M^{\alpha-3}\Big(\sum_{x^n\in\mathcal{T}_{\epsilon}^n(p_X)}\frac{p^{\alpha-3}(x^n,z^n)}{q^{\alpha-3}(z^n)}\Big)\Big(\sum_{x^n\in\mathcal{T}_{\epsilon}^n(p_X)}\frac{p^2(x^n,z^n)}{q^2(z^n)}\Big)+\cdots\Bigg]\nonumber\\
&\qquad+\cdots+(1+2\delta_n)^\alpha\label{type:osrb:tsallis:simplification:9},
\end{align}
\eqref{type:osrb:tsallis:simplification:6} arises from partitioning the set $\mathcal{X}^{\alpha}$ in a manner akin to \eqref{participating:nautral:numbersLproof} in the proof of Theorem \ref{RenyiOSRB}. Equation \eqref{type:osrb:tsallis:simplification:7} stems from the independent and uniform mapping of random binning.

\eqref{type:osrb:tsallis:simplification:9} is due to 
\begin{align}
\sum_{x_1^n\neq\cdots\neq x_{\ell}^n}\frac{\prod_{i=1}^{\ell}p^{\alpha_i}(x_i^n,z^n)}{q^{\alpha}(z^n)}\leq\prod_{i=1}^{\ell}\left(\sum_{x^n\in\mathcal{T}_{\epsilon}^n(p_X)}\frac{p^{\alpha_i}(x_i^n,z^n)}{q^{\alpha_i}(z^n)}\right),
\end{align}
and
\begin{align}
\frac{p(z^n)}{q(z^n)}=\sum_{x^n\in\mathcal{T}_{\epsilon}^n(p_X)}\frac{p(x^n,z^n)}{q(z^n)}=\frac{\sum_{x^n\in\mathcal{T}_{\epsilon}^n(p_X)}\tilde{p}(x^n)p(z^n\lvert x^n)}{\sum_{x^n\in\mathcal{T}_{\epsilon}^n(p_X)}q(x^n)p(z^n\lvert x^n)}\leq\frac{1}{1-\delta_n}\leq1+2\delta_n\label{OSRB:LD:bound:type:1}
\end{align}
with $q(z^n)=\prod_{i=1}^{n}p(z_i)$, and $\delta_n\to 0$ as $n\to\infty$.

For each realization of $Z^n = z^n$, let the probability measure $\nu_n(x)$ be defined as
\begin{align}
\nu_n(x^n)=\frac{p(x^n,z^n)}{\sum_{x^n\in\mathcal{T}_{\epsilon}^n(p_X)}p(x^n,z^n)}=\frac{p(x^n,z^n)}{p(z^n)}.
\end{align}
Upon defining $ f_i(x) = x^{\alpha_i - 1} $ and applying Harris' inequality, the terms in \eqref{type:osrb:tsallis:simplification:9} can be further bounded as follows.
\begin{align}
\prod_{i=1}^{\ell}\Bigg(\sum_{x^n\in\mathcal{T}_{\epsilon}^n(p_X)}\frac{p^{\alpha_i}(x^n,z^n)}{q^{\alpha_i}(z^n)}\Bigg)&= \Big(\frac{p(z^n)}{q(z^n)}\Big)^\ell \prod_{i=1}^{\ell}\mathbb{E}_{\nu_n(x^n)}\Big[f_i\Big(\frac{p(X^n,z^n)}{q(z^n)}\Big)\Big]\\
&\leq \Big(\frac{p(z^n)}{q(z^n)}\Big)^\ell\, \mathbb{E}_{\nu_n(x^n)}\left[\prod_{i=1}^{\ell}f_i\Big(\frac{p(X^n,z^n)}{q(z^n)}\Big)\right]\\
&=\Big(\frac{p(z^n)}{q(z^n)}\Big)^{\ell-1}\sum_{x^n\in\mathcal{T}_{\epsilon}^n(p_X)}\left[\frac{p(x^n,z^n)}{q(z^n)}\prod_{i=1}^{\ell}\left(\frac{p(x^n,z^n)}{q(z^n)}\right)^{\alpha_i-1}\right]\\
&\leq \left(1+2\delta_n\right)^{\ell-1}\sum_{x^n\in\mathcal{T}_{\epsilon}^n(p_X)}\left(\frac{p(x^n,z^n)}{q(z^n)}\right)^{\alpha-\ell+1}\!\!\!\!\!\!\!\!\!\!\!\!\!\!,\label{eqn:typic_sum}
\end{align} 
where \eqref{eqn:typic_sum} holds true because $\sum_{i=1}^{\ell}\alpha_i = \alpha$ and $\frac{p(z^n)}{q(z^n)} \leq 1 + 2\delta_n$. By substituting these upper bounds into \eqref{type:osrb:tsallis:simplification:9}, we obtain the following upper-bound terms for Tsallis divergence.
\begin{align}
\label{type:osrb:tsallis:simplification:10}
M^{\alpha-\ell}\sum_{z^n}q(z^n)\sum_{x^n\in\mathcal{T}_{\epsilon}^n(p_X)}\frac{p^{\alpha-\ell+1}(x^n,z^n)}{q^{\alpha-\ell+1}(z^n)}, \qquad \ell=1,\ldots,\alpha-1.
\end{align}
Supposing $X^n$ follows the distribution $\tilde{p}(x^n)$ within the set $\mathcal{T}_{\epsilon}^n\left(p_X\right)$ and $p_{Z^n\lvert X^n}=\prod_{i=1}^{n}p_{Z_i\lvert X_i}$, and assuming we uniformly and randomly partition the set $\mathcal{X}^n$ into $M=2^{nR}$ bins, Lemma \ref{asymptotic:analysis:on:rate:type} in Appendix \ref{appendix:lemma_D} suggests that as $n$ increases, each term in \eqref{type:osrb:tsallis:simplification:10} tends toward zero if the rate satisfies the subsequent inequality.
\begin{align}
R<H(X)-\sum_{x}p(x)D_{\alpha-\ell+1}\Big(p(z\lvert x)\parallel p(z)\Big),\qquad \ell=1,\ldots,\alpha-1,
\end{align}  
or equivalently
\begin{align}
R<\min_\ell H(X)&-{D}_{\alpha-\ell+1}\Big(p(z\lvert x)\parallel p(z)\Big)\overset{(a)}{=}H(X)-{D}_{\alpha}\Big(p(z\lvert x)\parallel p(z)\Big),
\end{align}
which $(a)$ results from the non-decreasing property of $D_{\alpha}\big(p(z\lvert x)\parallel p(z)\big)$.

Therefore, as $n$ increases, the upper bound \eqref{type:osrb:tsallis:simplification:9} tends toward $1$, and consequently, \eqref{type:osrb:tsallis:simplification:2} tends toward $0$. Hence, by letting $n$ approach infinity, the proof is completed.

\end{proof}

\begin{remark}
The converse proof of Theorem \ref{OSRB:type:achievability}, i.e., if $R > H(X) - \sum_{x} p(x) D_{\alpha}\big(p\left(z \lvert x\right) \parallel p\left(z\right)\big)$, then we have  
\begin{align}
\mathbb{E}_{\mathcal{B}}\left[T_\alpha\Big(P(b,z^n) \parallel p^{U}(b)\,q(z^n)\Big)\right] \to \infty,
\end{align} 
is similar to the converse proof of the i.i.d. case in Appendix \ref{converse:iid:case}, by repeating the steps \eqref{converse:1}-\eqref{converse:3}, which results in 
\begin{align}
\label{eqn:converse_type_ineq}
\mathbb{E}_{\mathcal{B}}\left[T_\alpha\Big(P(b,z^n) \parallel p^{U}(b)\,q(z^n)\Big)\right]\geq\frac{M^{\alpha-1}}{\alpha-1}\sum_{z^n,\,x^n\in\mathcal{T}_{\epsilon}^{n}(p_X)}q(z^n)\frac{p^{\alpha}(x^n, z^n)}{q^{\alpha}(z^n)}-\frac{1}{\alpha-1}.
\end{align}
The right-hand side of \eqref{eqn:converse_type_ineq} can be further lower-bounded to conclude the converse proof by changing the direction of the inequalities in Lemma \ref{asymptotic:analysis:on:rate:type}, replacing $H(p_X) - \delta_n(\epsilon)$ with $H(p_X) + \delta_n(\epsilon)$, and using the following inequalities instead of \eqref{simpification:binning:rate:type:1}--\eqref{simpification:binning:rate:type:2}:
\begin{align}
		&\log\left(\sum_{z^n,x^n\in\mathcal{T}_{\epsilon}^n(p_X)}q(z^n)\tilde{p}(x^n)\frac{p^{\alpha}(z^n\lvert x^n)}{q^{\alpha}(z^n)}\right)\nonumber\\
		&\qquad\qquad\geq\log\left(\left|\mathcal{T}_{\epsilon}^n(p_X)\right|\min_{x^n\in\mathcal{T}_{\epsilon}^n(p_X)}\sum_{z^n}q(z^n)\tilde{p}(x^n)\frac{p^{\alpha}(z^n\lvert x^n)}{q^{\alpha}(z^n)}\right)\\
		&\qquad\qquad\geq\log\left( 2^{-2n\delta_n(\epsilon)}\sum_{z^n}q(z^n)\frac{p^{\alpha}(z^n\lvert \bar{x}^n)}{q^{\alpha}(z^n)}\right)\\
		&\qquad\qquad\geq-2n\delta(\epsilon)+\log\prod_{x}\left(\sum_{z}p(z)\frac{p^{\alpha}(z\lvert{x})}{p^{\alpha}(z)}\right)^{n(p(x)-\epsilon)}\\
		&\qquad\qquad=\sum_{x}n(p(x)-2\delta_n(\epsilon)-\epsilon)\log\left(\sum_{z}p(z)\frac{p^{\alpha}(z\lvert x)}{p^{\alpha}(z)}\right).
	\end{align}

\end{remark}

\begin{remark}
	The achievability and converse proofs for $\alpha \in (0,1)$ in the special type case provide no gain, as their results coincide with those obtained for i.i.d. random variables for $\alpha \in (0,1)$. Therefore, they are omitted.
\end{remark}

 \begin{theorem}
 	\label{typical:inftyrandombinning}
 	
 	Assuming that $X^n \sim \tilde{p}(x^n)$ over $\mathcal{T}_{\epsilon}^n\left(p_X\right)$ and $p_{Z^n\lvert X^n}=\prod_{i=1}^{n}p_{Z_i\lvert X_i}$, if the binning rate meets the condition
\begin{align}
R<H(X)-\sum_{x}p(x)D_{\infty}\Big(p\left(z\lvert x\right)\parallel p\left(z\right)\Big),
\end{align}
then as $n$ approaches infinity, we find that
\begin{align}
\mathbb{E}_{\mathcal{B}}&\left[D_\infty\Big(P(b,z^n)\parallel p^{U}\!(b)\,q(z^n)\Big)\right]\to 0.
\end{align}
Here, $q(z^n)=\prod_{i=1}^{n}p(z_i)$ with $p(z)=\sum_{x}p(x)p(z\lvert x)$, and $H(X)=-\sum_{x}p(x)\log p(x)$.

 \end{theorem}
 
 \begin{proof}
The proof steps are similar to those in Theorem \ref{inftyrandombinning}, beginning in the asymptotic regime by considering sequences $x^n$ that belong to a special type of $\mathcal{T}_{\epsilon}^n(p_X)$. For any
 	\begin{align}
 	\label{eqn:beta_const}
 	 0<\beta\leq \dfrac{\exp\Big(n\left(H\left(X\right)-\delta_n\left(\epsilon\right)\right)-n\sum_{x}\left(p\left(x\right)+\epsilon\right)D_{\infty}\left(p\left(z\lvert x\right)\parallel p\left(z\right)\right)\Big)}{M(1+2\delta_n)},
 	\end{align}
 	we have
 	\begin{align}
 		\mathbb{E}_{\mathcal{B}}&\left[D_\infty\Big(P(b,z^n)\parallel p^{U}\!(b)\,q(z^n)\Big)\right]\label{typical:max:analysis:1}\\
 		&=\mathbb{E}_{\mathcal{B}}\left[\log\left(\max_{b,z^n}M\frac{P(b, z^n)}{q(z^n)}\right)\right]\label{typical:max:analysis:2}\\
 		&\leq\log\left(\mathbb{E}_{\mathcal{B}}\left[\max_{b,z^n}M\frac{P(b, z^n)}{q(z^n)}\right]\right)\label{typical:max:analysis:3}\\
 		&\leq\log\left(\frac{1}{\beta}\log\left(\sum_{b,z^n}\mathbb{E}_{\mathcal{B}}\left[\exp\left(\beta M\frac{P(b, z^n)}{q(z^n)}\right)\right]\right)\right)\label{typical:max:analysis:4}\\
 		&=\log\left(\frac{1}{\beta}\log\left(M\sum_{z^n}\mathbb{E}_\mathcal{B}\left[\exp\Big(\beta M\frac{P(1, z^n)}{q(z^n)}\Big)\right]\right)\right)\label{typical:max:analysis:5}\\
 		&=\log\left(\frac{1}{\beta}\log\left(M\sum_{z^n}\mathbb{E}_\mathcal{B}\left[\exp\Big(\beta M\,\frac{\sum_{x^n\in\mathcal{T}_{\epsilon}^n(p_X)}p(x^n, z^n)\mathds{1}\{\mathcal{B}(x^n)=1\}}{q(z^n)}\Big)\right]\right)\right)\\
 		&=\log\left(\frac{1}{\beta}\log\left(M\sum_{z^n}\mathbb{E}_\mathcal{B}\left[\exp\Big(\beta M \sum_{x^n\in\mathcal{T}_{\epsilon}^n(p_X)}\frac{p(x^n,z^n)}{q(z^n)}\mathds{1}\{\mathcal{B}(x^n)=1\}
 		\Big)\right]\right)\right)\\
 		&\leq\log\left(\frac{1}{\beta}\log\left(M\lvert\mathcal{Z}^n\rvert\max_{z^n}\left[\prod_{x^n\in\mathcal{T}_{\epsilon}^n(p_X)}\left(\frac{\exp\left(M\beta \frac{p(x^n, z^n)}{q(z^n)}\right)}{M}+1-\frac{1}{M}\right)\right]\right)\right)\label{typical:max:analysis:6}\\
 		&\leq\log\left(\frac{1}{\beta}\log\Bigg(M\rvert\mathcal{Z}^n\lvert\exp\left(\max_{z^n}\sum_{x^n\in\mathcal{T}_{\epsilon}^n(p_X)}\frac{1}{M}\left(\exp\left(M\beta \frac{p(x^n, z^n)}{q(z^n)}\right)-1\right)\right)\right)\label{typical:max:analysis:7}\\
 		&\leq\log\left(\frac{1}{\beta}\log \left(M\lvert\mathcal{Z}^n\rvert\right)+\max_{z^n}\Bigg[\sum_{x^n\in\mathcal{T}_{\epsilon}^n(p_X)}\frac{p(x^n,z^n)}{q(z^n)}+\beta M\sum_{x^n\in\mathcal{T}_{\epsilon}^n(p_X)} \frac{p^2(x^n,z^n)}{q^2(z^n)}\Bigg]\right)\label{typical:max:analysis:8}\\
 		&\leq\log\left(\frac{1}{\beta}\log \left(M\lvert\mathcal{Z}^n\rvert\right)+\max_{z^n}\left(\sum_{x^n\in\mathcal{T}_{\epsilon}^n(p_X)}\frac{p(x^n,z^n)}{q(z^n)}\right)\right.\label{typical:max:analysis:12}\\
 		&\left.\qquad\qquad\qquad\qquad\qquad+\beta M\left(\max_{x^n\in\mathcal{T}_{\epsilon}^n(p_X),\,z^n}\frac{p(x^n,z^n)}{q(z^n)}\right)\left(\sum_{x^n\in\mathcal{T}_{\epsilon}^n(p_X)} \frac{p(x^n,z^n)}{q(z^n)}\right)\right)\nonumber\\
 		&=\log\Bigg(\frac{1}{\beta}\log \left(M\lvert\mathcal{Z}^n\rvert\right)+\max_{z^n}\left(\frac{p(z^n)}{q(z^n)}\right)\label{typical:max:analysis:11}\\
 		&\qquad\qquad\qquad\qquad\qquad+\beta M\,\frac{p(z^n)}{q(z^n)}\max_{x^n\in\mathcal{T}_{\epsilon}^n(p_X),\,z^n}\left(\tilde{p}(x^n)\frac{p(z^n\lvert x^n)}{q(z^n)}\right)\Bigg)\nonumber\\
 		&\leq\log\Bigg(\frac{1}{\beta}\log\left(M\lvert\mathcal{Z}^n\rvert\right)+1+2\delta_n\label{typical:max:analysis:9}\\
 		&\qquad\qquad+\beta M(1+2\delta_n)\exp\Big(-n({H}(X)-\delta_n(\epsilon))+\max_{x^n\in\mathcal{T}_{\epsilon}^n(p_X),z^n}\frac{p(z^n\lvert x^n)}{q(z^n)}\Big)\Bigg)\nonumber\\
 		&\leq2\delta_n+\frac{n}{\beta}\log\lvert\mathcal{Z}\rvert+\frac{1}{\beta}\log M\label{typical:max:analysis:10}\\
 &\quad+M\beta(1+2\delta_n) \exp\Big(-n\left(H(X)-\delta_n(\epsilon)\right)+n\sum_{x}(p(x)+\epsilon)D_{\infty}\big(p(z\lvert x)\parallel p(z)\big)\Big).\nonumber
 	\end{align}
Here, \eqref{typical:max:analysis:3} arises from Jensen's inequality applied to $\log(x)$, while \eqref{typical:max:analysis:4} is derived from \eqref{eq:max_sum}, and \eqref{typical:max:analysis:5} follows from symmetry. Substituting \eqref{eqn:rand_bin_dist} into \eqref{typical:max:analysis:5}, and due to the independent and uniform mapping, we obtain \eqref{typical:max:analysis:6}. The result in \eqref{typical:max:analysis:7} comes from the inequality $1+x \leq \exp(x)$, and \eqref{typical:max:analysis:8} makes use of $e^x \leq 1 + x + x^2,, 0 \leq x \leq 1$, enforcing the following constraint on $\beta$
\begin{align}
\label{eqn:beta_prim_const}
0\leq \beta \leq \frac{1}{M\max\limits_{x^n\in\mathcal{T}_{\epsilon}^n(p_X),\,z^n}\frac{p(z^n\lvert x^n)}{q(z^n)}}.
\end{align}
\eqref{typical:max:analysis:9} uses \eqref{OSRB:LD:bound:type:1}, combined with
\begin{align}
\label{eqn:Typicality_p}
-H(X)-\delta_n(\epsilon)\leq\frac{1}{n}\log(\tilde{p}(x^n))\leq-H(X)+\delta_n(\epsilon),\qquad x^n\in\mathcal{T}_{\epsilon}^n(p_X),
\end{align}
resulting from \eqref{typical:distribution:defenition:1}. Finally, \eqref{typical:max:analysis:10} holds because $\log(1 + x) \leq x$, and
\begin{align}
\max_{x^n\in\mathcal{T}_{\epsilon}^n(p_X),z^n}\left(\frac{p(z^n\lvert x^n)}{q(z^n)}\right)&\leq\log\prod_{x}\left(\max_{z}\frac{p(z\lvert x)}{p(z)}\right)^{n(p(x)+\epsilon)}\\
&=n\sum_{x}(p(x)+\epsilon) D_{\infty}\big(p(z\lvert x)\parallel p(z)\big)\label{eqn:D_inf}.
\end{align}
Moreover, the constraint on $\beta$ in \eqref{eqn:beta_prim_const}, using \eqref{eqn:D_inf} and \eqref{eqn:Typicality_p}, can be considered as \eqref{eqn:beta_const}.

Taking the minimum of \eqref{typical:max:analysis:10} over $\beta$ in the interval given by \eqref{eqn:beta_const}, we obtain the following upper bound,
\begin{align}
&\mathbb{E}_{\mathcal{B}}\left[D_\infty\Big(p(b,z^n)\parallel p^{U}\!(b)\,q(z^n)\Big)\right]\nonumber\\
&\qquad\le\delta_n+ \sqrt{\frac{2(1+2\delta_n)M\log(M|\mathcal{Z}|^n)}{\exp\Big({n}({H}(X)-\delta_n(\epsilon))-n\sum_{x}(p(x)+\epsilon)D_{\infty}\big(p(z\lvert x)\parallel p(z)\big)\Big)}}\label{eqn:y0}
\end{align}
provided that
 	\begin{align}
 	 &\sqrt{\frac{(1+2\delta_n)\log(M|\mathcal{Z}|^n)}{M\exp\Big({n}({H}(X)-\delta_n(\epsilon))-n\sum_{x}(p(x)+\epsilon)D_{\infty}\big(p(z\lvert x)\parallel p(z)\big)\Big)}}\nonumber\\
 	 &\quad<\frac{1}{M(1+2\delta_n)}\exp\Big({n}({H}(X)-\delta_n(\epsilon))-n\sum_{x}(p(x)+\epsilon)D_{\infty}\big(p(z\lvert x)\parallel p(z)\big)\Big), 
 	  	\end{align}
 	 which is true if 
 	\begin{equation}\frac{M(1+2\delta_n)\log(M|\mathcal{Z}|^n)}{\exp\Big({n}({H}(X)+\delta_n(\epsilon))-n\sum_{x}(p(x)+\epsilon)D_{\infty}\big(p(z\lvert x)\parallel p(z)\big)\Big)}<1.\label{typical:eqn:y1}
 	\end{equation}
 
 Turning to the asymptotic regime, we consider $(X^n, Z^n)$ and set $M = 2^{nR}$. Then, if
\begin{align}
 	 R<\left({H}(X)-\delta_n(\epsilon)-\sum_{x}(p(x)+\epsilon)D_{\infty}\big(p(z\lvert x)\parallel p(z)\big)\right), 
\end{align}
the condition \eqref{typical:eqn:y1} is satisfied, and the upper bound in \eqref{eqn:y0} approaches zero as $n$ increases. This completes the proof.
 \end{proof}

\subsection{Random binning in the stochastic encoder}
Here, we consider the distribution of $(U^n, X^n)$ over the jointly $\epsilon$-typical set
\begin{align}
\mathcal{T}^n_{\epsilon}(p_{UX})=\{(u^n,x^n):u^n\in\mathcal{T}_{\epsilon}^n(p_U),\,x^n\in\mathcal{T}_{\epsilon}^n(X\lvert u^n)\},
\end{align}
where
\begin{align}
\mathcal{T}_{\epsilon}^n(X\lvert u^n)&=\left\{x^n:\Big\lvert\frac{1}{n}\sum_{i=1}^{n}\mathds{1}\{(x_i,u_i)=(x,u)\}-p(x,u)\Big\rvert<2\epsilon\right\},\\
\mathcal{T}_{\epsilon}^n(p_U)&=\left\{u^n:\Big\lvert\frac{1}{n}\sum_{i=1}^{n}\mathds{1}\{u_i=u\}-p(u)\Big\rvert<\epsilon\right\},
\end{align}
denoted as
\begin{align}
	\tilde{p}(u^n,x^n)=\frac{\prod_{i=1}^{n}p(u_i)p(x_i\lvert u_i)}{\left(\sum_{u^n\in\mathcal{T}_{\epsilon}^n(p_U)}\prod_{i=1}^{n}p(u_i)\right)\left(\sum_{x^n\in\mathcal{T}_{\epsilon}^n(X\lvert u^n)}\prod_{i=1}^{n}p(x_i\lvert u_i)\right)}.\label{stochastic:enc:osrb:3} 
\end{align}

	\begin{theorem}\label{Stocahstic:OSRB}
Assume the distribution of $(U^n, X^n,Z^n)$ is given by
\begin{align}
\tilde{p}(u^n,x^n,z^n)=\tilde{p}(u^n,x^n)\prod_{i=1}^{n}p(z_i\lvert x_i)\label{stochastic:enc:osrb:1} 
\end{align}
over all $(u^n,x^n)\in\mathcal{T}^n_{\epsilon}(p_{UX})$. Moreover, for $u^n \in \mathcal{T}_{\epsilon}^n(p_U)$, we have the random binning $\mathcal{B}(u^n):\,\mathcal{U}^n \to [2^{nR}]$. Let $\alpha \in (1,\infty)$. Defining 
\begin{align}
	&R_{\alpha}^\prime(p_X,p_{Z\lvert X},p_U)\label{auxiliary:random:u:alpha:finite}\\
	&\qquad=\!\!\!\!\max_{\substack{t_{Z\lvert XU}\\\sum_{x}p(x)p(z\lvert x)=p(z)\\\sum_{z}t(x\lvert u,z)t(z\lvert u)p(u)=p(u,x)}}\!\!\!\!\left[-\frac{\alpha}{\alpha-1}D\big(t(z\lvert u,x)\parallel p(z\lvert x)\lvert p(x,u)\big)+D\big(t(z\lvert u)\parallel p(z)\lvert p(u)\big)\right]\nonumber,
\end{align}
if the binning rate satisfies
 \begin{align}
	R<H(U)-R_{\alpha}^\prime(p_X,p_{Z\lvert X},p_U),\label{stochastic:enc:osrb:4} 
\end{align}
then, as $n$ goes to infinity, we obtain
\begin{align}
	\mathbb{E}_{\mathcal{B}}\left[T_{\alpha}\left(P(b,z^n)\parallel p^{U}(b)q(z^n)\right)\right]\to 0\label{stochastic:enc:osrb:5}
\end{align}
Here, we have $p(z)=\sum_{x}p(x)p(z\lvert x)$ and $q(z^n)=\prod_{i=1}^{n}p(z_i)$.
\end{theorem}

 \begin{proof}
 
\eqref{stochastic:enc:osrb:1}, along with the random binning of $\mathcal{U}^n$, implies that the joint distribution induced on the variables in this setup is as follows.
 	\begin{align}
 		P(b,u^n,x^n,z^n)&=\tilde{p}(u^n,x^n)\mathds{1}\{\mathcal{B}(u^n)=b\}\prod_{i=1}^{n}p(z_i\lvert x_i)\label{stochastic:enc:osrb:7}\\
 		P(b,z^n)&=\sum_{(u^n,x^n)\in\mathcal{T}^n_{\epsilon}(p_{UX})}\tilde{p}(u^n,x^n)\mathds{1}\{\mathcal{B}(u^n)=b\}\prod_{i=1}^{n}p(z_i\lvert x_i)\label{stochastic:enc:osrb:8}.
 	\end{align}
 	Here, we provide the proof for $\alpha \in \mathbb{N}$, and the proof for real $\alpha$ values, which follows a similar approach to the generalization proof of Theorem \ref{RenyiOSRB}, is omitted. Now, let us bound the Tsallis divergence, assuming $\alpha \in \mathbb{N}$.
 \begin{align}
 	&\mathbb{E}_{\mathcal{B}}\Big[(\alpha-1)T_{\alpha}\left(P(b,z^n)\parallel p^{U}(b)q(z^n)\right)\Big]\label{stochastic:enc:osrb:9}\\
 	&=M^{\alpha}\mathbb{E}_{\mathcal{B}}\left[\sum_{b,z^n}\frac{q(z^n)}{M}\left(\sum_{(u^n,x^n)\in\mathcal{T}_{\epsilon}^n(p_{UX})}\tilde{p}(u^n,x^n)\mathds{1}\{\mathcal{B}(u^n)=b\}\prod_{i=1}^{n}p(z_i\lvert x_i)\right)^{\alpha}q^{-\alpha}(z^n)\right]-1\label{stochastic:enc:osrb:10}\\
 	&=M^{\alpha}\mathbb{E}_{\mathcal{B}}\left[\sum_{z^n}q(z^n)\left(\sum_{(u^n,x^n)\in \mathcal{T}_{\epsilon}^n(p_{UX})}\tilde{p}(u^n,x^n)\mathds{1}\{\mathcal{B}(u^n)=1\}\prod_{i=1}^{n}p(z_i\lvert x_i)\right)^{\alpha}q^{-\alpha}(z^n)\right]-1\label{stochastic:enc:osrb:11}\\
 	&=M^{\alpha}\mathbb{E}_{\mathcal{B}}\left[\sum_{z^n}q(z^n)\left(\sum_{u^n\in\mathcal{T}_{\epsilon}^n(p_U)}\tilde{p}(u^n)S(z^n,u_i^n)\mathds{1}\{\mathcal{B}(u^n)=1\}\right)^\alpha q^{-\alpha}(z^n)\right]-1,\label{stochastic:enc:osrb:12}
 	\end{align}
where
 \begin{align}
 	S(z^n,u_i^n)\triangleq\sum_{x^n\in\mathcal{T}_{\epsilon}^n(X\lvert u_i^n)}\tilde{p}(x^n\lvert u_i^n)\prod_{i=1}^{n}p(z_i\lvert x_i).\label{stochastic:enc:osrb:13}
 \end{align}
Similar to the proof of Theorem \ref{RenyiOSRB}, \eqref{stochastic:enc:osrb:12} can be expanded as follows.
\begin{align}
 	&=M^{\alpha}\mathbb{E}_{\mathcal{B}}\Bigg[\sum_{z^n}q(z^n)\Bigg(\sum_{u_1^n\in\mathcal{T}_{\epsilon}^n(p_U)}\tilde{p}^{\alpha}(u_1^n)S^{\alpha}(z^n,u^n_1)q^{-\alpha}(z^n)\mathds{1}\{\mathcal{B}(u_1^n)=1\}\nonumber\\
 	&\quad+\sum_{(u^n_1\neq u^n_2)\in\mathcal{T}_{\epsilon}^n(p_U)}\tilde{p}^{\alpha-1}(u_1^n)\tilde{p}(u_2^n)S^{\alpha-1}(z^n,u_1^n)S(z^n,u_2^n)q^{-\alpha}(z^n)\prod_{i=1}^{2}\mathds{1}\{\mathcal{B}(u_i^n)=1\}+\cdots\nonumber\\
 	&\quad+\sum_{(u^n_1\neq u^n_2\neq u_3^n)\in\mathcal{T}_{\epsilon}^n(p_U)}\tilde{p}^{\alpha-2}(u_1^n)\tilde{p}(u^n_2)\tilde{p}(u^n_3)S^{\alpha-2}(z^n,u_1^n)S(z^n,u_2^n)S(z^n,u_3^n)q^{-\alpha}(z^n)\nonumber\\
&\qquad\qquad\qquad\qquad\qquad\qquad\qquad\qquad\qquad\qquad\qquad\qquad\qquad\times \prod_{i=1}^{3}\mathds{1}\{\mathcal{B}(u_i^n)=1\}+\cdots\nonumber\\
 	&\quad+\qquad\qquad\qquad\qquad\qquad\qquad\qquad\qquad\qquad\cdots\nonumber\\
 	&\quad+\sum_{(u^n_1\neq\cdots\neq u^n_{\alpha})\in\mathcal{T}_{\epsilon}^n(p_U)}\left(\prod_{i=1}^{\alpha}\tilde{p}(u_i^n)S(z^n,u^n_i)\mathds{1}\{\mathcal{B}(u_i^n)=1\}\right)q^{-\alpha}(z^n)\Bigg)\Bigg]-1\label{stochastic:enc:osrb:14}\\
 	&=M^{\alpha-1}\sum_{z^n}q(z^n)\Bigg(\sum_{u_1^n\in\mathcal{T}_{\epsilon}^n(p_U)}\frac{\tilde{p}^{\alpha}(u_1^n)S^{\alpha}(z^n,u^n_1)}{q^{\alpha}(z^n)}\Bigg)\nonumber\\
 	&\quad+M^{\alpha-2}\sum_{z^n}q(z^n)\left(\sum_{(u^n_1\neq u^n_2)\in\mathcal{T}_{\epsilon}^n(p_U)}\left(\frac{\tilde{p}^{\alpha-1}(u_1^n)S^{\alpha-1}(z^n,u_1^n)}{q^{\alpha-1}(z^n)}\right)\left(\frac{\tilde{p}(u_2^n)S(z^n,u_2^n)}{q(z^n)}\right)+\cdots\right)\nonumber\\
 	&\quad+M^{\alpha-3}\sum_{z^n}q(z^n)\Bigg(\sum_{(u^n_1\neq u^n_2\neq u_3^n)\in\mathcal{T}_{\epsilon}^n(p_U)}\left(\frac{\tilde{p}^{\alpha-2}(u_1^n)S^{\alpha-2}(z^n,u_1^n)}{q^{\alpha-2}(z^n)}\right)\left(\frac{\tilde{p}(u^n_2)S(z^n,u_2^n)}{q(z^n)}\right)\nonumber\\
 &\qquad\qquad\qquad\qquad\qquad\qquad\qquad\qquad\qquad\qquad\qquad\qquad\times\left(\frac{\tilde{p}(u^n_3)S(z^n,u_3^n)}{q(z^n)}\right)+\cdots\Bigg)\nonumber\\
    &\quad+\qquad\qquad\qquad\qquad\qquad\qquad\qquad\qquad\qquad\cdots\nonumber\\
 	&\quad+\sum_{z^n}q(z^n)\sum_{(u^n_1\neq\cdots\neq u^n_{\alpha})\in\mathcal{T}_{\epsilon}^n(p_U)}\prod_{i=1}^{\alpha}\left(\frac{\tilde{p}(u_i^n)S(z^n,u^n_i)}{q(z^n)}\right)-1\label{stochastic:enc:osrb:21}\\
 	&\leq M^{\alpha-1}\sum_{z^n}q(z^n)\left(\sum_{u^n\in\mathcal{T}_{\epsilon}^n(p_U)}\frac{\tilde{p}^{\alpha}(u^n)S^{\alpha}(z^n,u^n)}{q^{\alpha}(z^n)}\right)\nonumber\\
 	&\quad+M^{\alpha-2}\sum_{z^n}q(z^n)\left[\left(\sum_{u^n\in\mathcal{T}_{\epsilon}^n(p_U)}\frac{\tilde{p}^{\alpha-1}(u^n)S^{\alpha-1}(z^n,u^n)}{q^{\alpha-1}(z^n)}\right)\left(\sum_{u^n\in\mathcal{T}_{\epsilon}^n(p_U)}\frac{\tilde{p}(u^n)S(z^n,u^n)}{q(z^n)}\right)+\cdots\right]\nonumber\\
    &\quad+\qquad\qquad\qquad\qquad\qquad\qquad\qquad\qquad\qquad\cdots\nonumber\\
    &\quad+\sum_{z^n}q(z^n)\prod_{i=1}^{\alpha}\left[\left(\sum_{u^n\in\mathcal{T}_{\epsilon}^n(p_U)}\frac{\tilde{p}(u^n)S(z^n,u^n)}{q(z^n)}\right)\right]-1\label{stochastic:enc:osrb:24}\\
    &\leq M^{\alpha-1}\sum_{z^n}q(z^n)\left(\sum_{u^n\in\mathcal{T}_{\epsilon}^n(p_U)}\frac{\tilde{p}^{\alpha}(u^n)S^{\alpha}(z^n,u^n)}{q^{\alpha}(z^n)}\right)\nonumber\\
    &\quad+M^{\alpha-2}\sum_{z^n}q(z^n)\left[\left(\sum_{u^n\in\mathcal{T}_{\epsilon}^n(p_U)}\frac{\tilde{p}^{\alpha-1}(u^n)S^{\alpha-1}(z^n,u^n)}{q^{\alpha-1}(z^n)}\right)\left(1+2\delta_n\right)+\cdots\right]\nonumber\\
    &\quad+\qquad\qquad\qquad\qquad\qquad\qquad\qquad\qquad\qquad\cdots\nonumber\\
    &\quad+(1+2\delta_n)^{\alpha}-1.\label{stochastic:enc:osrb:27}
 \end{align}
\eqref{stochastic:enc:osrb:24} follows from 
\begin{align}
\sum_{(u_1\neq\cdots\neq u_{\ell})\in\mathcal{T}^n_{\epsilon}(p_U)}\prod_{i=1}^{\ell}\frac{\tilde{p}^{\alpha_i}(u_i^n)S^{\alpha_i}(z^n,u_i^n)}{q^{\alpha_i}(z^n)}\leq\prod_{i=1}^{\ell}\left(\sum_{u\in\mathcal{T}^n_{\epsilon}(p_U)}\frac{\tilde{p}^{\alpha_i}(u^n)S^{\alpha_i}(z^n,u^n)}{q^{\alpha_i}(z^n)}\right)\label{stochastic:enc:osrb:28}.
\end{align}
\eqref{stochastic:enc:osrb:27} is due to 
\begin{align}
	\sum_{u^n\in\mathcal{T}_{\epsilon}^n(p_U)}\frac{\tilde{p}(u^n)S(z^n,u^n)}{q(z^n)}&=\frac{\sum_{(u^n,x^n)\in\mathcal{T}_{\epsilon}^{n}(p_{UX})}\tilde{p}(u^n,x^n)\prod_{i=1}^{n}p(z_i\lvert x_i)}{q(z^n)}\label{stochastic:enc:osrb:29}\\
	&=\frac{\sum_{(u^n,x^n)\in\mathcal{T}_{\epsilon}^{n}(p_{UX})}\tilde{p}(u^n,x^n)\prod_{i=1}^{n}p(z_i\lvert x_i)}{\sum_{x^n,u^n}\prod_{i=1}^{n}p(u_i,x_i)\prod_{i=1}^{n}p(z_i\lvert x_i)}\label{stochastic:enc:osrb:30}\\
	&\leq\frac{\sum_{(u^n,x^n)\in\mathcal{T}_{\epsilon}^{n}(p_{UX})}\tilde{p}(u^n,x^n)\prod_{i=1}^{n}p(z_i\lvert x_i)}{\sum_{(x^n,u^n)\in\mathcal{T}_{\epsilon}^{n}(p_{UX})}\prod_{i=1}^{n}p(u_i,x_i)\prod_{i=1}^{n}p(z_i\lvert x_i)}\label{stochastic:enc:osrb:31}\\
	&\leq\frac{1}{1-\delta_n}\leq 1+2\delta_n\label{stochastic:enc:osrb:32},
\end{align}

where \eqref{stochastic:enc:osrb:32} follows from \eqref{stochastic:enc:osrb:3}, knowing that
\begin{align}
\sum_{x^n\in\mathcal{T}_{\epsilon}^n(p_{UX})}\prod_{i=1}^{n}p(u_i, x_i)\geq 1-\delta_n
\end{align}
with $\delta_n$ approaching zero as $n$ becomes large.

Let $Z^n = z^n$ be fixed, and define the distribution $\nu(u^n)$ over $u^n\in\mathcal{T}_{\epsilon}^n(p_U)$ as follows.
\begin{align}
\nu(u^n)=\frac{\frac{\tilde{p}(u^n)S(u^n,z^n)}{q(z^n)}}{\sum_{u^n\in\mathcal{T}_{\epsilon}^n(p_U)}\frac{\tilde{p}(u^n)S(u^n,z^n)}{q(z^n)}}=\frac{\tilde{p}(u^n)S(u^n,z^n)}{\sum_{u^n\in\mathcal{T}_{\epsilon}^n(p_U)}\tilde{p}(u^n)S(u^n,z^n)}.\label{stochastic:enc:osrb:33}
\end{align}
Then, using Harris' inequality for the non-decreasing function $f_i(x) = x^{\alpha_i - 1}$ with $\sum_{i=1}^{\ell} \alpha_i = \alpha$, we obtain
\begin{align}
&\prod_{i=1}^{\ell}\left[\sum_{u^n\in\mathcal{T}_{\epsilon}^n(p_U)}\left(\frac{\tilde{p}(u^n)S(z^n,u^n)}{q(z^n)}\right)^{\alpha_i}\right]\nonumber\\
&= \left(\frac{q(z^n)}{\sum_{u^n\in\mathcal{T}^n_{\epsilon}(p_U)}\tilde{p}(u^n)S(z^n,u^n)}\right)^{-\ell}\prod_{i=1}^{\ell}\mathbb{E}_{\nu(u^n)}\left[f_i\left(\frac{\tilde{p}(U^n)S(U^n,z^n)}{q(z^n)}\right)\right]\\
&\leq \left(\frac{q(z^n)}{\sum_{u^n\in\mathcal{T}^n_{\epsilon}(p_U)}\tilde{p}(u^n)S(z^n,u^n)}\right)^{-\ell}\mathbb{E}_{\nu(u^n)}\left[\prod_{i=1}^{\ell}f_i\left(\frac{\tilde{p}(U^n)S(U^n,z^n)}{q(z^n)}\right)\right]\\
&=\left(\frac{q(z^n)}{\sum\limits_{u^n\in\mathcal{T}^n_{\epsilon}(p_U)}\tilde{p}(u^n)S(z^n,u^n)}\right)^{1-\ell}\left[\sum_{u^n\in\mathcal{T}^n_{\epsilon}(p_U)}\frac{\tilde{p}(u^n)S(u^n,z^n)}{q(z^n)}\prod_{i=1}^{\ell}\left(\frac{\tilde{p}(u^n)S(u^n,z^n)}{q(z^n)}\right)^{\alpha_i-1}\right]\\
&\leq\left(1+2\delta_n\right)^{\ell-1}\left[\sum_{u^n\in\mathcal{T}^n_{\epsilon}(p_U)}\frac{\tilde{p}(u^n)S(u^n,z^n)}{q(z^n)}\prod_{i=1}^{\ell}\left(\frac{\tilde{p}(u^n)S(u^n,z^n)}{q(z^n)}\right)^{\alpha_i-1}\right]\label{eqn:typicality_q}\\
&=\left(1+2\delta_n\right)^{\ell-1}\sum_{u^n\in\mathcal{T}^n_{\epsilon}(p_U)}\left(\frac{\tilde{p}(u^n)S(u^n,z^n)}{q(z^n)}\right)^{\alpha-\ell+1},\label{eqn:upper_bound_typ_Har}
\end{align}
where \eqref{eqn:typicality_q} follows from \eqref{stochastic:enc:osrb:32}. Using the upper bound in \eqref{eqn:upper_bound_typ_Har} to bound the terms in \eqref{stochastic:enc:osrb:27}, we obtain the following terms.
\begin{align}
\label{eqn:upper_Typ_R}
\left(1+2\delta_n\right)^{\ell-1}M^{\alpha-\ell}\sum_{z^n}q(z^n)\sum_{u^n\in\mathcal{T}^n_{\epsilon}(p_U)}\left(\frac{\tilde{p}(u^n)S(u^n,z^n)}{q(z^n)}\right)^{\alpha-\ell+1}.
\end{align}
Setting $M = 2^{nR}$ in \eqref{eqn:upper_Typ_R} suggests that, as $n$ increases, each term in \eqref{stochastic:enc:osrb:27} tends toward zero if the rate $R$ satisfies the following inequality.
\begin{align}
	R<\min_{\ell\in[\alpha-1]}\lim_{n\to\infty}\frac{-1}{n(\alpha-\ell)}\log\left(\sum_{z^n}q(z^n)\sum_{u^n\in\mathcal{T}_{\epsilon}^n(p_U)}\left(\frac{\tilde{p}^{\alpha-\ell+1}(u^n)S^{\alpha-\ell+1}(u^n,z^n)}{q^{\alpha-\ell+1}(z^n)}\right)\right)\label{stochastic:enc:osrb:39}.
\end{align}
Here, the rate constraint can be further simplified as follows.
\begin{align}
	&\lim_{n\to\infty}\frac{1}{n(\alpha-\ell)}\log\left({\sum_{z^n}q(z^n)\sum_{u^n\in\mathcal{T}_{\epsilon}^n(p_U)}\left(\frac{\tilde{p}^{\alpha-\ell+1}(u^n)S^{\alpha-\ell+1}(u^n,z^n)}{q^{\alpha-\ell+1}(z^n)}\right)}\right)\label{stochastic:asymptotic:analysis:on:rate:type:4}\\
	&=\lim_{n\to\infty}\frac{1}{n(\alpha-\ell)}\log\left({\sum_{z^n,u^n\in\mathcal{T}_{\epsilon}^n(p_U)}q(z^n)\left(\frac{\tilde{p}^{\alpha-\ell+1}(u^n)S^{\alpha-\ell+1}(u^n,z^n)}{q^{\alpha-\ell+1}(z^n)}\right)}\right)\label{stochastic:asymptotic:analysis:on:rate:type:5}\\
	&\leq\lim_{n\to\infty}\frac{1}{n(\alpha-\ell)}\Bigg(\log\left[\max_{u^n\in\mathcal{T}_{\epsilon}(p_U)} \tilde{p}^{\alpha-\ell}(u^n)\right]\nonumber\\
	&\qquad\qquad\qquad\qquad\qquad\qquad+\log\left[{\sum_{z^n,u^n\in\mathcal{T}_{\epsilon}^n(p_U)}q(z^n)\left(\frac{\tilde{p}(u^n)S^{\alpha-\ell+1}(u^n,z^n)}{q^{\alpha-\ell+1}(z^n)}\right)}\right]\Bigg)\label{stochastic:asymptotic:analysis:on:rate:type:6}
	\\
	&\leq\lim_{n\to\infty} \Bigg[-(H(U)-\delta^\prime_n(\epsilon))\nonumber\\
	&\quad\qquad+\frac{1}{n(\alpha-\ell)}\log\left((1+2\delta_n(\epsilon))^{\alpha-\ell}{\sum_{z^n,u^n\in\mathcal{T}_{\epsilon}^n(p_U)}q(z^n)\left(\frac{\tilde{p}(u^n)S^{\alpha-\ell+1}(u^n,z^n)}{q^{\alpha-\ell+1}(z^n)}\right)}\right)\Bigg]\label{stochastic:asymptotic:analysis:on:rate:type:7}\\
	&=-H(U)+\lim_{n\to\infty}\frac{1}{n(\alpha-\ell)}\log\left({\sum_{z^n,u^n\in\mathcal{T}_{\epsilon}^n(p_U)}}\tilde{p}(u^n)S^{\alpha-\ell+1}(u^n,z^n)q^{\ell-\alpha}(z^n)\right)\label{stochastic:asymptotic:analysis:on:rate:type:8}\\
	&\leq-H(U)\nonumber\\
	&+\max_{\substack{t_{Z\lvert XU}\\\sum_{x}p(x)p(z\lvert x)=p(z)\\\sum_{z}t(x\lvert u,z)t(z\lvert u)p(u)=p(u,x)}}\left[-\frac{\alpha-\ell+1}{\alpha-\ell}D\Big(t(z\lvert u,x)\parallel p(z\lvert x)\big\lvert p(x,u)\Big)+D\Big(t(z\lvert u)\parallel p(z)\big\lvert p(u)\Big)\right]\label{stochastic:asymptotic:analysis:on:rate:type:9}
\end{align}
\eqref{stochastic:asymptotic:analysis:on:rate:type:7} follows from 
\begin{align}
2^{-n(H(U)+\delta^\prime_n(\epsilon))}\leq\prod_{i=1}^{n}p(u_i)\leq\tilde{p}(u^n)&=\frac{\prod_{i=1}^{n}p(u_i)}{\sum_{u^n\in\mathcal{T}^n_{\epsilon}(p_{U})}\prod_{i=1}^{n}p(u_i)}\label{eqn:left_type_u}\\
&\leq \frac{1}{1-\delta_n(\epsilon)}\prod_{i=1}^{n}p(u_i)\\
&\leq (1+2\delta_n(\epsilon))\prod_{i=1}^{n}p(u_i)\\
&\leq (1+2\delta_n(\epsilon))2^{-n(H(U)-\delta^\prime_n(\epsilon))}\label{eqn:right_type_u}.
\end{align}
The upper bound in \eqref{stochastic:asymptotic:analysis:on:rate:type:9}, with $s = \alpha - \ell$ and $w^n = u^n$, is computed in \cite[(450)–(455)]{yu2018renyi}. It is straightforward to check that setting $\ell = 1$ maximizes this upper bound, which completes the proof.

 \end{proof}

\begin{remark}

The converse part can be obtained similarly to Theorem \ref{OSRB:type:achievability} using \eqref{eqn:left_type_u}--\eqref{eqn:right_type_u}, combined with
\begin{align}
	\sum_{u^n\in \mathcal{T}^n_{\epsilon}(p_{U})}\prod_{i=1}^{n}p(u_i)\geq 1-\delta_n(\epsilon).
\end{align}

\end{remark}

\begin{remark}
	The achievability and converse proofs for a stochastic encoder in the case of $\alpha \in (0, 1)$ follow similar steps to Theorem \ref{Theorem:OSRB:alpha:(0,1)}, resulting in the condition $R < H(U \lvert Z)$ on the random binning rate, instead of $H(X \lvert Z)$.
\end{remark}

\begin{remark}
	Assuming $D_{\infty}$, and following similar steps as in the proof of Theorem \ref{typical:inftyrandombinning}, we obtain the following binning rate constraint.
\begin{align}
		R<H(U)-R_{\infty}^\prime(p_X,p_{Z\lvert X},p_U)\label{stochastic:enc:osrb:41},
\end{align}
where
\begin{align}
&R_{\infty}^\prime(p_X,p_{Z\lvert X},p_U)\nonumber\\
&\quad=\max_{\substack{t_{Z\lvert XU}\\\sum_{x}p(x)p(z\lvert x)=p(z)\\\sum_{z}t(x\lvert u,z)t(z\lvert u)p(u)=p(u,x)}}\Big[-D\big(t(z\lvert u,x)\parallel p(z\lvert x)\big\lvert p(x,u)\big)+D\big(t(z\lvert u)\parallel p(z)\big\lvert p(u)\big)\Big]\label{auxiliary:random:u:alpha:infinite}.
\end{align}
\end{remark}

\section{Applications}
\label{sec:applications}

In this section, we explore the application of Tsallis-based OSRB in analyzing the achievability rate regions for the wiretap channel problem.

The wiretap channel is modeled as a conditional distribution $p_{YZ\lvert X}$, so that $X$ is the channel's input, $Y$ is the information received by the legitimate receiver, and $Z$ is the leakage output received by the eavesdropper. Assuming the message set is $\mathcal{M}=[2^{nR}]$, the goal is to send the maximum amount of information securely to the legal receiver. Considering mutual information as a measure of security, it is well known that the rate $R<I(X;Y)-I(X;Z)$ can be achieved for any input distribution $p(x)$. Here, we assume Tsallis divergence as a security measure and provide an asymptotic analysis of the achievable rate.

\subsection{Wiretap channel with deterministic encoder}
\begin{theorem}
\label{thm:asymp_wiretap}
Let $p_{YZ \lvert X}$ be an arbitrary wiretap channel. Additionally, assume $q(z^n) = \prod_{i=1}^{n} p(z_i)$ with $p(z) = \sum_{x} p(x) p(z \lvert x)$, $\mathcal{M} = [2^{nR}]$ as the message set, and $\alpha \in (1, \infty)$. Then, for the following distribution over $\mathcal{T}_{\epsilon}^n(p_X)$,
\begin{align}
	\tilde{p}(x^n)=\frac{\prod_{i=1}^{n}p(x_i)}{\sum_{x^n\in\mathcal{T}_{\epsilon}^n(p_X)}\prod_{i=1}^{n}p(x_i)},
\end{align}
there exists a code that can send a message $M$ with rate $R$, reliably with vanishing error and strongly securely 
\begin{align}
\lim_{n\to\infty} T_\alpha\Big(p(m,z^n)\parallel p^{U}\!(m)\,q(z^n)\Big)=0.
\end{align}
Provided that the rate satisfies
\begin{align}
\label{eqn:secure_rate_alpha}
R<I(X;Y)-\sum_{x}p(x)D_{\alpha}\Big(p(z\lvert x)\parallel p(z)\Big).
\end{align}
\end{theorem}

\begin{proof}
Let $\beta=(m,f):X^n\to[2^{nR_{1}}]\times[2^{nR_2}]$ be a random binning that uniformly and independently maps each sequence $x^n$ to two bin indices, $m\in[2^{nR_1}]$ and $f\in[2^{nR_2}]$. More precisely, if the condition $R_1+R_2<H(X)$ holds \cite[Theorem 1]{yassaee2014achievability}, then we have
\begin{align}
\label{eqn:indep_cond_mes}
\lim_{n\rightarrow\infty}\mathbb{E}_\mathcal{B}\left[\|P(m,f)-p^{U}\!(m)\,p^{U}\!(f)\|_{\mathrm{TV}}\right]=0
\end{align}
In addition, if the condition $R_2>H(X|Y)$ is met, the decoder, having $(F,Y^n)$, can reliably decode the sequence $X^n$. More specifically, we have
\begin{align}
\label{eqn:SW_cond}
\lim_{n\rightarrow\infty}\mathbb{E}_\mathcal{B}\left[\mathbb{P}\left(\hat{X}^n\neq X^n\right)\right]=0,
\end{align}
by taking the expectation over all random binnings $\mathcal{B}$. Finally, using Theorem \ref{RenyiOSRB}, the constraint $R_1+R_2<H(X)-\sum_{x}p(x)D_{\alpha}\big(p(z\lvert x)\parallel p(z)\big)$ results that
\begin{align}
\label{leakgeexp1}
\lim_{n\to\infty} \mathbb{E}_\mathcal{B}\left[T_\alpha\Big(P(m,f,z^n)\parallel p^{U}\!(m)\,p^{U}\!(f)\,q(z^n)\Big)\right]=0.
\end{align}
On the other hand, using Lemma \ref{lemma:D_cond} in Appendix \ref{subsec:useful_lemma}, we obtain that for at least half of the $(\mathcal{B},F)$ pairs
\begin{align}
\label{leakgeexp2}
\lim_{n\to\infty} \,T_\alpha\Big(p(m,z^n|f)\parallel p^{U}\!(m)\,q(z^n)\Big)=0.
\end{align}
Therefore, by choosing the positive real numbers $R_1,R_2$ so that
\begin{align}
&R_1+R_2< H(X)\label{eq:cond_indep}\\
&R_2>H(X|Y)\\
&R_1+R_2<H(X)-\sum_{x}p(x)D_{\alpha}\Big(p(z\lvert x)\parallel p(z)\Big)\label{eq:cond_secrecy},
\end{align}
the reliability and security constraints are established, and $M$ and $F$ become independent. Based on property \ref{prop:second} of Lemma \ref{decreasinglemma}, it can be concluded that $H(X)-\sum_{x}p(x)D_{\alpha}\big(p(z\lvert x)\parallel p(z)\big)\leq H(X|Z)\leq H(X)$, which shows that satisfying condition \eqref{eq:cond_secrecy} leads to satisfying condition \eqref{eq:cond_indep}. Therefore, condition \eqref{eq:cond_indep} can be ignored.
 
Therefore, using \eqref{eqn:SW_cond}, \eqref{leakgeexp2} and Markov's inequality together with a union bound, for any 
$\epsilon^\prime>0$ and sufficiently large $n$, there exists $f\in[2^{nR_2}]$ and a random labeling $\beta_0$ such that
\begin{align}\label{errorvanish}
\mathbb{P}\left[\hat{X}^n\neq X^n\big\lvert\beta_0, F = f\right]\leq\epsilon^\prime,
\end{align}
and
\begin{align}
\label{negligibleleakagefix}
T_\alpha\Big(p(m,z^n|F=f)\parallel p^{U}\!(m)\,q(z^n)\Big)\rightarrow 0.
\end{align}
Now we construct the code as follows. We consider $M$ as the message that is uniformly distributed according to \eqref{eqn:indep_cond_mes}, and then we transmit the codeword $X^n=\beta_0^{-1}(M,f)$. Based on \eqref{errorvanish}, the legitimate receiver decodes the message with an asymptotically vanishing error and the eavesdropper does not obtain any information from the message according to \eqref{negligibleleakagefix}.

Take note that our coding strategy establishes the distribution as 
\begin{align}
p(m|f)\,p(x^n|m,f)\,p(z^n|x^n).
\end{align}
However, it is imperative that condition \eqref{negligibleleakagefix} is satisfied with the ensuing distribution. 
\begin{align}
\label{eqn:main_prob}
p^U(m)\,p(x^n|m,f)\,p(z^n|x^n).
\end{align}
This assurance is provided by Lemma \ref{lemma:tsallis_mess}, which guarantees that a vanishing decrease in message rate results in an asymptotically diminishing Tsallis divergence for the distribution in \eqref{eqn:main_prob}. This concludes the proof.
\end{proof}

\begin{corollary}
For $\alpha=1$, the rate $R=I(X;Y)-I(X;Z)$ can be achieved with an asymptotically vanishing error and with a strong security measure of $I(M;Z^n)\leq\epsilon$.
\end{corollary}

\begin{proof}
The proof according to property \ref{prop:alpha1} of Lemma \ref{decreasinglemma} is straightforward.
\end{proof}

\begin{remark}
Inequality \eqref{Tisalis:D:Compare1} suggests that all the achievability proofs for $T_{\alpha}$ as a security measure also holds for $D_{\alpha}$ when $\alpha\in(1,\infty)$.
\end{remark}

\begin{remark}
Suppose we are in the $\alpha\in(0,1)$ case. Using Theorem \ref{Theorem:OSRB:alpha:(0,1)} and similar to the proof of Theorem \ref{thm:asymp_wiretap}, the achievable secrecy rate with secrecy criterion $T_\alpha$ is equal to $R<H(X|Z)-H(X|Y)$. In \cite{yu2018renyi}, taking into account the stronger criterion $D_\alpha$, the same achievable region has been obtained.
\end{remark}

As already mentioned, due to the infinity of $T_\infty$, we consider $D_\infty$ instead. Therefore, using Theorem \ref{typical:inftyrandombinning}, we have the following secure rate region.
\begin{theorem}
\label{thm:asymp_wiretap_infty}
For an arbitrary wiretap channel, assume $q(z^n) = \prod_{i=1}^{n} p(z_i)$ with $p(z) = \sum_{x} p(x) p(z \lvert x)$. Then, there exists a code.
 that can send a message $M$ with rate $R$, reliably with vanishing error and strongly securely with 
\begin{align}
\lim_{n\to\infty} D_\infty\Big(p(m,z^n)\parallel p^{U}\!(m)\,q(z^n)\Big)=0.
\end{align}
Provided that the rate satisfies
\begin{align}
\label{eqn:secure_rate_infty}
R<I(X;Y)-\sum_{x}p(x)D_{\infty}\Big(p(z\lvert x)\parallel p(z)\Big).
\end{align}
\end{theorem}
\begin{proof}
The proof steps are similar to those in Theorem \ref{thm:asymp_wiretap}. We choose positive real numbers $R_1$ and $R_2$ such that
\begin{align}
R&=R_1\\
R_2&>H(X\lvert Y)\label{eqn:e_error}\\
R_1+R_2&<H(X)-\sum_{x}p(x)D_{\infty}\Big(p(z\lvert x)\parallel p(z)\Big).\label{eqn:delta_secure}
\end{align}
\eqref{eqn:e_error} guarantees negligible-error decodability, and based on Theorem \ref{typical:inftyrandombinning}, condition \eqref{eqn:delta_secure} ensures strong security. Eliminating $R_2$ from \eqref{eqn:e_error} and \eqref{eqn:delta_secure} and setting the message rate equal to $R_1$ completes the proof of the theorem.
\end{proof}

\begin{corollary}

	If the binned sequences are generated i.i.d. instead of choosing among typical sequences, then by repeating all the steps in Theorem \ref{thm:asymp_wiretap} and Theorem \ref{thm:asymp_wiretap_infty} with Theorem \ref{RenyiOSRB} and Theorem \ref{inftyrandombinning}, we obtain the achievable secure rates as
\begin{align}
R<\tilde{H}_{\alpha}(X\lvert Z)-H(X\lvert Y)\label{comparing:iid:typicalset:10}
\end{align} 
and
\begin{align}
R<\tilde{H}_{\infty}(X\lvert Z)-H(X\lvert Y),\label{comparing:iid:typicalset:11}
\end{align}
which are respectively weaker than the results obtained in \eqref{eqn:secure_rate_alpha} and \eqref{eqn:secure_rate_infty}.

\end{corollary}

\begin{proof}

To show that \eqref{comparing:iid:typicalset:10} is a weaker result compared to \eqref{eqn:secure_rate_alpha}, we have
\begin{align}
    \tilde{H}_{\alpha}(X \lvert Z) \leq H(X) - \sum_{x} p(x) D_{\alpha}\Big(p(z \lvert x) \parallel p(z)\Big), \label{comparing:iid:typicalset:12}
\end{align}
which can be proven by following these steps:
\begin{align}
H(X)&-\sum_{x}p(x)D_{\alpha}\Big(p(z\lvert x)\parallel p(z)\Big)\label{comparing:iid:typicalset:1}\\
&=\sum_{x}p(x)\log\frac{1}{p(x)}-\frac{1}{\alpha-1}\sum_{x}p(x)\log\left(\sum_{z}p(z)\left(\frac{p(z\lvert x)}{p(z)}\right)^{\alpha}\right)\label{comparing:iid:typicalset:2}\\
&=\frac{1}{\alpha-1}\sum_{x}p(x)\log\left(\frac{1}{p(x)}\right)^{\alpha-1}-\frac{1}{\alpha-1}\sum_{x}p(x)\log\left(\sum_{z}p(z)\left(\frac{p(z\lvert x)}{p(z)}\right)^{\alpha}\right)\label{comparing:iid:typicalset:3}\\
&=\frac{1}{\alpha-1}\sum_{x}p(x)\log\left(\frac{1}{p^{\alpha-1}(x)\sum_{z}p(z)\left(\frac{p(z\lvert x)}{p(z)}\right)^{\alpha}}\right)\label{comparing:iid:typicalset:4}\\
&=\frac{1}{1-\alpha}\left(\sum_{x}p(x)\log\left(\sum_{z}p^{\alpha-1}(x)p^{1-\alpha}(z)p^{\alpha}(z\lvert x)\right)\right)\label{comparing:iid:typicalset:5}\\
&\geq\frac{1}{1-\alpha}\left(\log\left(\sum_{x,z}p^{\alpha}(x)p^{1-\alpha}(z)p^{\alpha}(z\lvert x)\right)\right)\label{comparing:iid:typicalset:6}\\
&=\frac{1}{1-\alpha}\left(\log\sum_{x,z}p(z)\left(\frac{p(x,z)}{p(z)}\right)^{\alpha}\right)\label{comparing:iid:typicalset:7}\\
&=\frac{1}{1-\alpha}\log\left(\sum_{z}p(z)\sum_{x}p^{\alpha}(x\lvert z)\right)\label{comparing:iid:typicalset:8}\\
&=\tilde{H}_{\alpha}(X\lvert Z),\label{comparing:iid:typicalset:9}
\end{align}
where \eqref{comparing:iid:typicalset:6} follows from Jensen's inequality for the function $f(x) = \log x$.

To demonstrate that \eqref{comparing:iid:typicalset:11} represents a weaker result compared to \eqref{eqn:secure_rate_infty}, we need to show that
\begin{align}
\tilde{H}_{\infty}(X\lvert Z)\leq H(X)-\sum_{x}p(x)D_{\infty}\Big(p(z\lvert x )\parallel p(z)\Big),\label{comparing:iid:typicalset:13}
\end{align}
which can be established through the following steps:
\begin{align}
	H(X)&-\sum_{x}p(x)D_{\infty}\Big(p(z\lvert x)\parallel p(z)\Big)\\
	&=\sum_{x}p(x)\log\frac{1}{p(x)}-\sum_{x}p(x)\log\left(\max_{z}\frac{p(z\lvert x)}{p(z)}\right)\\
	&=\sum_{x}p(x)\log\left(\frac{1}{p(x)\left(\max_{z}\frac{p(z\lvert x)}{p(z)}\right)}\right)\\
	&=\sum_{x}p(x)\log\left(\frac{1}{\max_{z}\frac{p(x,z)}{p(z)}}\right)\\
	&\geq\log\left(\frac{1}{\max_{x,z}\frac{p(x,z)}{p(z)}}\right)\\
	&=\tilde{H}_{\infty}(X\lvert Z).
\end{align}
\end{proof}

\subsection{Wiretap channel with stochastic encoder}
In this section, we derive an optimal rate for a wiretap channel when the encoder is stochastic rather than deterministic, using the results obtained in Theorem \ref{Stocahstic:OSRB}.

\begin{theorem}\label{secure:stochastic:encoder}
Let $p_{YZ \lvert X}$ be an arbitrary wiretap channel. Assume the joint distribution $p(u,x)$ with $U \sim p(u)$ such that $\lvert \mathcal{U} \rvert \leq \lvert \mathcal{X} \rvert + 1$. Then, there exists a code satisfying the rate constraint
\begin{align}
    R < I(U, Y) - R_{\alpha}^\prime(p_X, p_{Z \lvert X}, p_U),
\end{align}
which ensures that the message can be decoded with vanishing error while simultaneously satisfying the strong secrecy constraint:
\begin{align}
    \lim_{n \to \infty} T_{\alpha}\left(p(m, z^n) \parallel p^{U}(m) q(z^n)\right) = 0,
\end{align}
with $q(z^n) = \prod_{i=1}^{n} p(z_i)$ and $p(z) = \sum_{x} p(x) p(z \lvert x)$, where $R_{\alpha}^\prime(p_X, p_{Z \lvert X}, p_U)$ is defined in \eqref{auxiliary:random:u:alpha:finite}.

\end{theorem}

\begin{proof}
The proof structure is similar to that of the wiretap channel with a deterministic encoder.

Consider an arbitrary random variable $U^n$ over $\mathcal{T}_{\epsilon}^n(p_U)$ with distribution
\begin{align}
    \tilde{p}(u^n) = \frac{\prod_{i=1}^{n} p(u_i)}{\sum_{u^n \in \mathcal{T}^n_{\epsilon}(p_{U})} \prod_{i=1}^{n} p(u_i)}.
\end{align}
Let $\beta = (m, f): U^n \to [2^{nR_1}] \times [2^{nR_2}]$ be a random binning that maps each sequence $u^n$ uniformly and independently to bin indices. Then, the binning rate must satisfy certain constraints to ensure both secrecy and reliability.
\begin{itemize}
	\item Secrecy condition: Based on Theorem \ref{Stocahstic:OSRB}, the secrecy condition can be satisfied if
\begin{align}
    & R = R_1, \\
    & R_1 + R_2 < H(U) - R_{\alpha}^\prime(p_X, p_{Z \lvert X}, p_U). \label{wiretap:OSRB:stochastic:2}
\end{align}  
More precisely, under condition \eqref{wiretap:OSRB:stochastic:2}, we have
\begin{align}
\lim_{n\to\infty}\mathbb{E}_{\mathcal{B}}\left[T_{\alpha}\left(p(m,f,z^n)\parallel p^{U}(m)p^{U}(f)p(z^n)\right)\right]=0.\label{wiretap:OSRB:stochastic:3}
\end{align}
\item Reliability condition: To satisfy the reliability condition, consider $R_2 > H(U \lvert Y)$. Therefore, given $(F, Y^n)$, the sequence $U^n$ can be decoded with vanishing error, or equivalently,
\begin{align}
	\lim_{n\to\infty}\mathbb{E}_{\mathcal{B}}\left[\mathbb{P}\left[\hat{U}^n\neq U^n\right]\right]=0,\label{wiretap:OSRB:stochastic:4}
\end{align} 
where $\mathbb{E}_{\mathcal{B}}$ denotes the expectation over all random mappings.
\end{itemize}

Lemma \ref{lemma:D_cond} implies that there exists a fixed realization of $(\mathcal{B}, \mathcal{F})$ such that
\begin{align}
&\lim_{n\to\infty}T_{\alpha}\left(p(m,z^n\lvert f)\parallel p^{U}(m)q(z^n)\right)=0\label{wiretap:OSRB:stochastic:5}\\
&\qquad\lim_{n\to\infty}\mathbb{P}\left[\hat{U}^n\neq U^n\big\lvert\beta_0, F = f\right]=0\label{wiretap:OSRB:stochastic:6}.
\end{align} 
Our coding strategy establishes the distribution as 
\begin{align}
    p(m \lvert f) \, \tilde{p}(u^n \lvert m, f) \, \tilde{p}(x^n \lvert u^n) \, p(z^n \lvert x^n).
\end{align}
However, we require that condition \eqref{wiretap:OSRB:stochastic:5} is satisfied with the following distribution:
\begin{align}
    \label{eqn:main_prob:stochastic}
    p^U(m) \, \tilde{p}(u^n \lvert m, f) \, \tilde{p}(x^n \lvert u^n) \, p(z^n \lvert x^n).
\end{align}
This requirement is ensured by Lemma \ref{lemma:tsallis_mess} and Remark \ref{stochstic:uniform:messages}, which guarantee that a vanishing decrease in message rate results in an asymptotically diminishing Tsallis divergence for the distribution in \eqref{eqn:main_prob:stochastic}, concluding the proof.
\end{proof}

\begin{remark}
	The proof of the secrecy rate under the $D_{\infty}$ secrecy constraint for a stochastic encoder follows similar steps, with rates given by
\begin{align}
    & R = R_1, \\
    & R_1 + R_2 < H(U) - R_{\infty}^\prime(p_X, p_{Z \lvert X}, p_U), \\
    & R_2 > H(U \lvert Y),
\end{align}
where $R_{\infty}^\prime(p_X, p_{Z \lvert X}, p_U)$ is defined in \eqref{auxiliary:random:u:alpha:infinite}.
\end{remark}

	\begin{remark}
The achievable secure rate using the OSRB method for $\alpha \in (0, 1)$ under Tsallis divergence is given by $I(U; Y) - I(U; Z)$, which is similar to the mutual information criterion and thus is omitted.
	\end{remark}

	\begin{remark}
		In \cite{yu2018renyi}, the secure rate is computed for $\alpha \in (0, 2] \cup \{\infty\}$ for both deterministic and stochastic encoders. The secure rates computed in Theorem \ref{thm:asymp_wiretap} and Theorem \ref{secure:stochastic:encoder} for $\alpha \in (0, \infty) \cup \{\infty\}$ coincide with the results in \cite[Theorems 5-6]{yu2018renyi} for $\alpha \in (0, 2]\cup\{\infty\}$ and extend the findings to $\alpha \in (2, \infty)$.
	\end{remark}

\begin{remark}
 $T_\alpha$ and $D_\alpha$ exhibit similar behavior for $\alpha \in (0, \infty)$ as $n \to \infty$, demonstrating that the secure capacity with $D_\alpha$ is equal to that with $T_\alpha$ for $\alpha \in (0, \infty)$. The achievable rates under the $T_\alpha$ secrecy constraint in Theorem \eqref{thm:asymp_wiretap} and Theorem \eqref{secure:stochastic:encoder}, which are equal to one as presented in \cite{yu2018renyi} with $D_\alpha$ (which has also been proven to represent the capacity), imply that the Tsallis-based secure rates computed in this paper are also the capacity.
\end{remark}

\section{Conclusion}
\label{sec:conclusion}

In this paper, we conducted an asymptotic analysis of OSRB using the Tsallis divergence criterion. Specifically, we examined the conditions on the binning rate, denoted as $R$, that would satisfy the following expectation over all random binnings: \begin{align} 
\lim_{n \to \infty} \mathbb{E}_\mathcal{B}\left[T_\alpha\Big(p(m, z^n) \parallel p^{U}(m) q(z^n)\Big)\right] = 0, 
\end{align} 
where $X$ and $Z$ are random variables with a joint distribution $p(x, z)$, and $\beta: \mathcal{X}^n \to [M]$ is a random binning function. Our analysis covers the entire range of $\alpha \in (0, \infty) \cup \{\infty\}$, provides both achievability and converse proofs, and considers three scenarios: (i) the binned sequence is generated i.i.d., (ii) the sequence is randomly chosen from an $\epsilon$-typical set, and (iii) the sequence originates from an $\epsilon$-typical set and is passed through a non-memoryless virtual channel. Additionally, we investigated the binning rate by employing $D_\infty$ as an alternative to $T_\infty$. By leveraging the established theorems for asymptotic OSRB analysis, we were able to examine the achievable rate region for the wiretap channel.

\bibliographystyle{IEEEtran}
\bibliography{Reference}

\appendix

\subsection{Proof of Lemma \ref{decreasinglemma}}
\label{subsec:app_prop}

\begin{enumerate}
\item Let $X$ and $Z$ be two independent random variables or equivalently $p(x,z)=p(x)p(z)$, hence
\begin{align}
\tilde{H}_{\alpha}(X\lvert Z)&=\frac{1}{1-\alpha}\log\left(\sum_{z}p(z)\sum_{x}p^{\alpha}(x|z)\right)\\
&=\frac{1}{1-\alpha}\log\left(\sum_{z}p(z)\sum_{x}p^{\alpha}(x)\right)\\
&=\frac{1}{1-\alpha}\log\left(\sum_{x}p^{\alpha}(x)\right)\\
&=H_\alpha(X).
\end{align}

\item To prove it, it suffices to show that $\frac{d\tilde{H}_{\alpha}(X|Z)}{d\alpha}<0$ holds for $\alpha\in(1,\infty)$.
\begin{align}
\frac{d\tilde{H}_{\alpha}(X|Z)}{d\alpha}&=\frac{1}{(\alpha-1)^2}\log\left(\sum_{z}p(z)\sum_{x}p^{\alpha}(x|z)\right)\nonumber\\
&\qquad+\frac{1}{1-\alpha}\frac{\sum_{z}p(z)\sum_{x}p^{\alpha}(x|z)\log p(x|z)}{\sum_{z}p(z)\sum_{x}p^{\alpha}(x|z)}\\
&=\frac{1}{(\alpha-1)^2}\log\Bigg(\sum_{z}p(z)\sum_{x}p^\alpha(x|z)\Bigg)\nonumber\\
&\qquad-\frac{1}{(\alpha-1)^2}\frac{\sum_{z}p(z)\sum_{x}p^{\alpha}(x|z)\log p^{\alpha-1}_{x\lvert z}}{\sum_{z}p(z)\sum_{x}p^\alpha(x|z)}\\
&=\frac{\mathbb{E}\Big[p^{\alpha-1}_{x\lvert z}\Big]\log\mathbb{E}\Big[p^{\alpha-1}_{x\lvert z}\Big]-\mathbb{E}\Big[p^{\alpha-1}_{x\lvert z}\log p^{\alpha-1}_{x\lvert z}\Big]}{(\alpha-1)^2\,\mathbb{E}\Big[p^{\alpha-1}_{x\lvert z}\Big]}\label{eqn:expec_def}\\
&\leq0.\label{Jensendecreasement}
\end{align}
\eqref{eqn:expec_def} is obtained by taking the common denominator and the following 
\begin{align}
\mathbb{E}\left[p^{\alpha-1}_{x\lvert z}\right]&=\sum_{x,z}p(x,z)\, p(x|z)^{\alpha-1}\\
&=\sum_{z}p(z)\sum_{x}p^\alpha(x|z),
\end{align}
where the expectation is over $p(x,z)$. Moreover, \eqref{Jensendecreasement} follows from Jensen's inequality for the convex function $g(c)=c\log c$. Further, ``$=$" occurs when the channel is singleton.

\item Using L'H\^{o}pital's rule, we get
\begin{align}
\lim_{\alpha\to 1}\tilde{H}_{\alpha}(X\lvert Z)&=-\frac{d}{d\alpha}\left(\log\sum_{z}p(z)\sum_{x}p^{\alpha}_{x\lvert z}\right)\Bigg\lvert_{\alpha=1}\\
&=-\frac{\sum_{z}p(z)\sum_{x}p^{\alpha}(x|z)\log p(x|z)}{\sum_{z}p(z)\sum_{x}p^{\alpha}(x|z)}\Bigg\lvert_{\alpha=1}\\
&=-\mathbb{E}_{X,Z}\Big[\log p(x|z)\Big]\\
&=H(X\lvert Z).
\end{align}

\item Data processing inequality
\begin{align}
\tilde{H}_{\alpha}(X|Y)&=\frac{1}{1-\alpha}\log\left(\sum_{y}p(y)\sum_{x}p^{\alpha}(x|y)\right)\\
&=\frac{1}{1-\alpha}\log\left(\mathbb{E}_{Y}\Big[\sum_{x}p^{\alpha}(x|Y)\Big]\right)\\
&=\frac{1}{1-\alpha}\log\left(\mathbb{E}_{Y,Z}\left[\sum_{x}p^{\alpha}(x|Y)\right]\right)\\
&=\frac{1}{1-\alpha}\log\left(\mathbb{E}_{Z}\mathbb{E}_{Y|Z}\left[\sum_{x}p^{\alpha}(x|Y)\right]\right)\\
&\leq\frac{1}{1-\alpha}\log\left(\mathbb{E}_{Z}\left[\sum_{x}p^{\alpha}(x|Z)\right]\right)\label{Jensenfordataproc}\\
&=\tilde{H}_{\alpha}(X|Z).
\end{align}
\eqref{Jensenfordataproc} follows from Jensen's inequality for $g(c)=c^\alpha,\alpha>1$ and $p_{x|Z}=\mathbb{E}_{Y|Z}\left[p(x|Y)\right]$.
\item The proof is straightforward using Jensen's inequality for the concave function $g(c)=\log c$.

\item By tending $\alpha$ to infinity and keeping the maximum term, the result is
\begin{align}
\lim_{\alpha\to\infty}\tilde{H}_{\alpha}(X\lvert Z)&=\lim_{\alpha\to\infty}\frac{1}{1-\alpha}\log\left(\sum_{z,x}p(z)p^{\alpha}(x\lvert z)\right)\\
&=\lim_{\alpha\to\infty}\frac{1}{1-\alpha}\log\left(p(z)\cdot\big(\max_{x,z}p(x|z)\big)^\alpha\right)\\
&=\lim_{\alpha\to\infty}\frac{1}{1-\alpha}\log p(z)\nonumber\\
&\qquad+\lim_{\alpha\to\infty}\frac{\alpha}{1-\alpha}\log\left(\max_{x,z}p(x|z)\right)\\
&=\log\left(\frac{1}{\max_{x,z}p(x|z)}\right).
\end{align}

\item Let $p(x^n,z^n)=\prod_{i=1}^{n}p(x_i,z_i)$ then
\begin{align}
\tilde{H}_{\alpha}(X^n\lvert Z^n)&=\frac{1}{1-\alpha}\log\Bigg(\sum_{z^n}p(z^n)\sum_{x^n}p^{\alpha}(x^n\lvert z^n)\Bigg)\\
&=\frac{1}{1-\alpha}\log\left(\sum_{z^n}\prod_{i=1}^{n}p(z_i)\sum_{x^n}\prod_{i=1}^{n}p^{\alpha}(x_i\lvert z_i)\right)\\
&=\frac{1}{1-\alpha}\log\prod_{i=1}^{n}\Bigg(\sum_{z_i}p(z_i)\sum_{x_i}p^{\alpha}(x_i\lvert z_i)\Bigg)\\
&=\sum_{i=1}^n\tilde{H}_{\alpha}(X_i\lvert Z_i)\\
&=n\tilde{H}_{\alpha}(X\lvert Z).
\end{align}

\item According to the property of decreasing $\tilde{H}_{\alpha}(X\lvert Z)$ with respect to $\alpha$ (property \ref{prop:second}) and non-negativity of $\tilde{H}_{\infty}(X\lvert Z)$, the proof is complete.
\begin{align}
\label{simple1}
\tilde{H}_{\alpha}(X\lvert Z)\geq\tilde{H}_{\infty}(X\lvert Z)=\log\left(\frac{1}{\max_{x,z}p(x|z)}\right)\geq0.
\end{align}

\item According to the definition of the singleton channel, we have
\begin{align}
\max_{x,z} p(x,z)=\min_{x,z} p(x,z),
\end{align}
which subsequently, based on property \ref{prop:infty}, leads to
\begin{align}
\tilde{H}_{\infty}(X\lvert Z)=\tilde{H}_{-\infty}(X\lvert Z).
\end{align}
The decreasing property of $\tilde{H}_{\alpha}(X\lvert Z)$ completes the proof.
\end{enumerate}

\subsection{Proof of Theorem \ref{RenyiOSRB} for real \texorpdfstring{$\alpha$'s}{}}
\label{sec:app_real}

\underline{Achievability part of the proof}\\
Starting from \eqref{expand1} for real $\alpha\in(1,\infty)$ and decomposing $\alpha=\floor{\alpha}+\fr{\alpha}$ into its integer and decimal parts, we have
\begin{align}
&M^{\alpha}\mathbb{E}_{\mathcal{B}}\left[\left(\sum_{x}p(x\lvert z)\mathds{1}\{\mathcal{B}(x)=1\}\right)^{\alpha}\right]\nonumber\\
&=M^{\alpha}\mathbb{E}_{\mathcal{B}}\left[\left(\sum_{x}p(x\lvert z)\mathds{1}\{\mathcal{B}(x)=1\}\right)^{\floor{\alpha}}\left(\sum_{\bar{x}}p(\bar{x}\lvert z)\mathds{1}\{\mathcal{B}(\bar{x})=1\}\right)^{\fr{\alpha}}\right]\label{real:simplification:2}\\
&\leq M^{\alpha}\sum_{\alpha_1+\cdots+\alpha_\ell=\floor{\alpha}}\left(\sum_{x_1\neq\cdots\neq x_\ell}\mathbb{E}_{\mathcal{B}(x_1),\ldots,\mathcal{B}(x_\ell)}\left[\prod_{i=1}^{\ell}p^{\alpha_i}(x_i\lvert z)\mathds{1}\{\mathcal{B}(x_i)=1\}\right]\right)\label{real:simplification:3}\\
&\qquad\qquad\qquad\qquad\qquad\cdot\left(\mathbb{E}_{\mathcal{B}(\bar{x})\lvert\mathcal{B}(x_1),\ldots,\mathcal{B}(x_\ell)}\left[\sum_{\bar{x}}p(\bar{x}\lvert z)\mathds{1}\{\mathcal{B}(\bar{x})=1\}\right]\right)^{\fr{\alpha}}\label{real:simplification:4}\\
&\leq M^{\alpha-\ell}\sum_{\alpha_1+\cdots+\alpha_\ell=\floor{\alpha}}\sum_{x_1\neq\cdots\neq x_\ell}\left(\prod_{i=1}^{\ell}p^{\alpha_i}(x_i\lvert z)\right)\cdot\left(\frac{1}{M}+\sum_{i=1}^{\ell}p(x_i\lvert z)\right)^{\fr{\alpha}}\label{real:simplification:5}\\
&\leq M^{\alpha-\ell}\sum_{\alpha_1+\cdots+\alpha_\ell=\floor{\alpha}}\sum_{x_1\neq\cdots\neq x_\ell}\left(\prod_{i=1}^{\ell}p^{\alpha_i}(x_i\lvert z)\right)\cdot\left(M^{\floor{\alpha}-\alpha}+\sum_{i=1}^{\ell}p^{\fr{\alpha}}(x_i\lvert z)\right)\label{real:simplification:6}\\
&=\sum_{\alpha_1+\cdots+\alpha_\ell=\floor{\alpha}}\Bigg(M^{\floor\alpha-\ell}\sum_{x_1\neq\cdots\neq x_\ell}\prod_{i=1}^{\ell}p^{\alpha_i}(x_i\lvert z)\nonumber\\
&\qquad\qquad\qquad\qquad\qquad+\sum_{j=1}^{\ell}M^{\alpha-\ell}\sum_{x_1\neq\cdots\neq x_\ell}\bigg(p^{\alpha_j+\fr{\alpha}}(x_j\lvert z)\prod_{\substack{i=1\\i\neq j}}^{\ell}p^{\alpha_i}(x_i\lvert z)\bigg)\Bigg)\label{real:simplification:7}\\
&\leq\sum_{\alpha_1+\cdots+\alpha_\ell=\floor{\alpha}}\Bigg(M^{\floor\alpha-\ell}\bigg(\prod_{i=1}^{\ell}\sum_{x_i}p^{\alpha_i}(x_i\lvert z)\bigg)\nonumber\\
&\qquad\qquad\qquad\qquad\qquad+\sum_{j=1}^{\ell}M^{\alpha-\ell}\bigg(\prod_{\substack{i=1\\i\neq j}}^{\ell}\sum_{x_i}p^{\alpha_i}(x_i\lvert z)\bigg)\cdot\bigg(\sum_{x_j}p^{\alpha_j+\fr{\alpha}}(x_j\lvert z)\bigg)\Bigg)\label{real:simplification:8},
\end{align}
where \eqref{real:simplification:3} is the result of Jensen's inequality for the concave function $f(x)=x^{\fr{\alpha}}$. Inequality \eqref{real:simplification:6} is a result of the following for non-negative $x_i$'s.
\begin{align}
\left(\sum_{i}x_i\right)^{\fr{\alpha}}\leq \sum_{i}x_i^{\fr{\alpha}}.
\end{align}
Moreover, \eqref{real:simplification:8} comes from
\begin{align}
\sum_{x_1\neq x_2\neq\cdots\neq x_{\ell}}\prod_{i=1}^{\ell}p^{\alpha_i}(x_i\lvert z)\leq\prod_{i=1}^{\ell}\sum_{x_i}p^{\alpha_i}(x_i\lvert z).
\end{align} 
The terms in \eqref{real:simplification:8} can be further upper bounded by Harris inequality. By fixing $Z = z$, knowing that $\sum_{i=1}^{\ell}\alpha_{i}=\floor\alpha$, and using the nondecreasing function $f_i(x) = x^{\alpha_i - 1}$, we have:
\begin{align}
\prod_{i=1}^{\ell}\sum_{x}p^{\alpha_i}(x\lvert z)&=\prod_{i=1}^{\ell}\mathbb{E}_{p_{X\lvert Z=z}}\left[f_i\Big(p(X\lvert z)\Big)\right]\nonumber\\
&\leq \mathbb{E}_{p_{X\lvert Z=z}}\left[\prod_{i=1}^{\ell}f_i\Big(p(X\lvert z)\Big)\right]\nonumber\\
&=\mathbb{E}_{p_{X\lvert Z=z}}\left[p^{\floor\alpha-\ell}(X\lvert z)\right]=\sum_{x}p^{\floor\alpha-\ell+1}(x\lvert z).\label{eq:harris1}
\end{align}
By applying Harris inequality once more and considering $f_i(x)=x^{\alpha_{i}-1}$ and $g_j(x)=x^{\alpha_{j}+\{\alpha\}-1}$, we obtain
\begin{align}
\Big(\sum_{x}p^{a_j+\{\alpha\}}(x\lvert z)\Big)\prod_{\substack{i=1\\i\neq j}}^{\ell}\Bigg[\sum_{x}p^{\alpha_i}(x\lvert z)\Bigg]&=\mathbb{E}_{p_{X\lvert Z=z}}\Bigg[g_j(p(X\lvert z))\Bigg]\prod_{\substack{i=1\\i\neq j}}^{\ell}\mathbb{E}_{p_{X\lvert Z=z}}\Bigg[f_i(p(X\lvert z))\Bigg]\nonumber\\
&\leq \mathbb{E}_{p_{X\lvert Z=z}}\Bigg[g_j(p(X\lvert z))\prod_{\substack{i=1\\i\neq j}}^{\ell}f_i(p(X\lvert z))\Bigg]\nonumber\\
&=\mathbb{E}_{p_{X\lvert Z=z}}\Bigg[p^{\alpha-\ell}(X\lvert z)\Bigg]=\sum_{x}p^{\alpha-\ell+1}(x\lvert z).\label{eq:harris2}
 \end{align}
Considering the upper bounds in \eqref{eq:harris1} and \eqref{eq:harris2}, substituting \eqref{real:simplification:8} into \eqref{eq:expec}, and taking into account $(X^n, Z^n)$ and $M = [2^{nR}]$, the rate $R$ must satisfy the following inequality:
\begin{align}
R<\min_{\substack{i\in[\ell]\\1\leq \alpha_i\leq\floor\alpha}}\left\{\tilde{H}_{\alpha_i}(X\lvert Z),\tilde{H}_{\alpha_i+\fr{\alpha}}(X\lvert Z)\right\}=\tilde{H}_\alpha(X|Z),
\end{align} 
where equality results from the decreasing property of $\tilde{H}_\alpha(X|Z)$ with $\alpha$. This completes the achievability proof.

\underline{Converse part of the proof}\label{converse:iid:case}\\
Our objective is to demonstrate that for $\alpha\in(1,\infty)$, if $R>\tilde{H}_{\alpha}(X\lvert Z)$, then it follows that
\begin{align}
\mathbb{E}_{\mathcal{B}}\left[T_\alpha\left(P(b,z^n)\parallel p^{U}(b)p(z^n)\right)\right]\rightarrow\infty.
\end{align}

By putting \eqref{expand1} in \eqref{eq:expec}, we have
\begin{align}
\mathbb{E}_{\mathcal{B}}&\left[T_\alpha\Big(P(b,z)\parallel p^{U}\!(b)\,p(z)\Big)\right]\nonumber\\
&=\frac{1}{\alpha-1}\left(M^{\alpha}\mathbb{E}_{Z}\mathbb{E}_{\mathcal{B}}\left[\left(\sum_{x}p(x\lvert Z)\mathds{1}\{\mathcal{B}(x)=1\}\right)^{\alpha}\right]-1\right)\label{converse:1}\\
&\geq \frac{M^{\alpha}}{\alpha-1}\mathbb{E}_{Z}\mathbb{E}_{\mathcal{B}}\left[\sum_{x}p^{\alpha}(x\lvert Z)\mathds{1}\{\mathcal{B}(x)=1\}\right]-\frac{1}{\alpha-1}\label{converse:2}\\
&=\frac{M^{\alpha-1}}{\alpha-1}\sum_{z}p(z)\sum_{x}p^{\alpha}(x\lvert z)-\frac{1}{\alpha-1}\label{converse:3}.
\end{align}
\eqref{converse:2} is the result of the following inequality for $\alpha>1$.
\begin{align}
\left(\sum_{i}x_i\right)^{\alpha}\geq\sum_{i}x^{\alpha}_i.
\end{align}
Let $(X^n,Z^n)$ and $M=[2^{nR}]$, then we have
\begin{align}
\mathbb{E}_{\mathcal{B}}\left[T_\alpha\Big(P(b,z^n)\parallel p^{U}\!(b)\,p(z^n)\Big)\right]\geq 2^{(\alpha-1)n(R-\tilde{H}_\alpha(X|Z))}-\frac{1}{\alpha-1},
\end{align}
which with $R>\tilde{H}_\alpha(X|Z)$, the right side tends to infinity as $n$ increases, and the converse part of the proof is complete.

\subsection{Proof of Theorem \ref{Theorem:OSRB:alpha:(0,1)}}\label{Proof:OSRB:alpha:(0,1)}

\begin{proof}\

\underline{Achievability part of the proof}
\begin{align}
&(\alpha-1)\cdot\mathbb{E}_{\mathcal{B}}\left[T_\alpha\Big(P(b,z)\parallel p^{U}\!(b)\,p(z)\Big)\right]\nonumber\\
&=\mathbb{E}_\mathcal{B}\left[\sum_{b,z}\frac{p(z)}{M}\cdot\Big(M^{\alpha}P^{\,\alpha}(b\lvert z)-1\Big)\right]\\
&=\mathbb{E}_\mathcal{B}\left[\sum_{z}p(z)\Big(M^{\alpha}P^{\,\alpha}(b=1\lvert z)-1\Big)\right]\label{Tisalis:(0,1):3}\\
&=\mathbb{E}_Z\mathbb{E}_{\mathcal{B}}\left[M^{\alpha}\left(\sum_{x}p(x\lvert z)\mathds{1}\{\mathcal{B}(x)=1\}\right)^{\alpha}-1\right]\label{Tisalis:(0,1):4}\\
&=\mathbb{E}_Z\mathbb{E}_{\mathcal{B}}\left[M^{\alpha}\frac{\sum_{x}p(x\lvert z)\mathds{1}\{\mathcal{B}(x)=1\}}{\Big(\sum_{\bar{x}}p(\bar{x}\lvert z)\mathds{1}\{\mathcal{B}(\bar{x})=1\}\Big)^{1-\alpha}}-1\right]\label{Tisalis:(0,1):5}\\
&=\mathbb{E}_Z\mathbb{E}_{\mathcal{B}}\left[M^{\alpha}\left(\sum_{x}p(x\lvert z)\frac{\mathds{1}\{\mathcal{B}(x)=1\}}{\left(\sum_{\bar{x}}p(\bar{x}\lvert z)\mathds{1}\{\mathcal{B}(\bar{x})=1\}\right)^{1-\alpha}}\right)-1\right]\label{Tisalis:(0,1):6}\\
&=\mathbb{E}_Z\mathbb{E}_{\mathcal{B}(x)}\mathbb{E}_{\mathcal{B}(\bar{x})\lvert \mathcal{B}(x)}\left[M^{\alpha}\left(\sum_{x}p(x\lvert z)\frac{\mathds{1}\{\mathcal{B}(x)=1\}}{\left(\sum_{\bar{x}}p(\bar{x}\lvert z)\mathds{1}\{\mathcal{B}(\bar{x})=1\}\right)^{1-\alpha}}\right)-1\right]\label{Tisalis:(0,1):7}\\
&\geq\mathbb{E}_Z\mathbb{E}_{\mathcal{B}(x)}\left[M^{\alpha}\left(\sum_{x}p(x\lvert z)\frac{\mathds{1}\{\mathcal{B}(x)=1\}}{\mathbb{E}_{\mathcal{B}(\bar{x})\lvert \mathcal{B}(x)}\left[\Big(\sum_{\bar{x}}p(\bar{x}\lvert z)\mathds{1}\{\mathcal{B}(\bar{x})=1\}\Big)^{1-\alpha}\right]}\right)-1\right]\label{Tisalis:(0,1):8}\\
&\geq\mathbb{E}_Z\mathbb{E}_{\mathcal{B}(x)}\left[M^{\alpha}\left(\sum_{x}p(x\lvert z)\frac{\mathds{1}\{\mathcal{B}(x)=1\}}{\bigg(\mathbb{E}_{\mathcal{B}(\bar{x})\lvert \mathcal{B}(x)}\Big[\sum_{\bar{x}}p(\bar{x}\lvert z)\mathds{1}\{\mathcal{B}(\bar{x})=1\}\Big]\bigg)^{1-\alpha}}\right)-1\right]\label{Tisalis:(0,1):9}\\
&\geq\mathbb{E}_Z\left[M^{\alpha}\left(\sum_{x}p(x\lvert z)\frac{\frac{1}{M}}{\Big(\frac{1}{M}+p(x\lvert z)\Big)^{1-\alpha}}\right)-1\right]\label{Tisalis:(0,1):10}\\
&=\mathbb{E}_{X,Z}\left[\left(\frac{1}{1+Mp(x\lvert z)}\right)^{{1-\alpha}}-1\right]\label{Tisalis:(0,1):11},
\end{align}
where \eqref{Tisalis:(0,1):8} and \eqref{Tisalis:(0,1):9} are obtained by applying Jensen's inequality on functions $f(x)=\frac{1}{x}$ and $g(x)=x^{1-\alpha}$, respectively.

Let $M=[2^{nR}]$ and $(X^n,Z^n)$, it suffices to prove that by increasing $n$ 
\begin{align}
\mathbb{E}_{X,Z}\left[\left(\frac{1}{1+Mp(x\lvert z)}\right)^{1-\alpha}\right]\to 1.
\end{align}
Using \eqref{Tisalis:(0,1):11}, We define the following typical set.
\begin{align}
\mathcal{A}^{n}_{\epsilon}=\{(x^n,z^n):\frac{1}{n}h(x^n\lvert z^n)\geq H(X\lvert Z)-\epsilon\},
\end{align}
 which $h(x\lvert z)=\log\frac{1}{p(x\lvert z)}$, and $\epsilon$ is an arbitrary positive number. By the weak law of large numbers, we have $\lim_{n\to\infty}p(\mathcal{A}^{n}_{\epsilon})=1$, so we can write
\begin{align}
\mathbb{E}_{X,Z}\left[\left(\frac{1}{1+2^{n(R-h(X^n\lvert Z^n))}}\right)^{1-\alpha}\right]
&\geq\mathbb{E}_{X,Z}\left[\left(\frac{1}{1+2^{n(R-h(X^n\lvert Z^n))}}\right)^{1-\alpha}\mathds{1}\{(x^n,z^n)\in\mathcal{A}^{n}_{\epsilon}\}\right]\\
&\geq\mathbb{E}_{X,Z}\left[\left(\frac{1}{1+2^{n\left(R-H(X\lvert Z)+\epsilon\right)}}\right)^{1-\alpha}\mathds{1}\{(x^n,z^n)\in\mathcal{A}^{n}_{\epsilon}\}\right]\\&=p(\mathcal{A}^{n}_{\epsilon})\left[\left(\frac{1}{1+2^{n\left(R-H(X\lvert Z)+\epsilon\right)}}\right)^{1-\alpha}\right]\\
&\to\left[\left(\frac{1}{1+2^{n\left(R-H(X\lvert Z)+\epsilon\right)}}\right)^{1-\alpha}\right]\label{simplification:tisalis:(0,1)}.
\end{align}
Therefore, with the increase of $n$, the condition $R<H(X\lvert Z)-\epsilon$ makes \eqref{simplification:tisalis:(0,1)} tend to $1$ and the achievability proof is complete.

\underline{Converse part of the proof}

Starting from the definition of $T_{\alpha}$ in \eqref{eqn:tsallis_divergence}, we have
\begin{align}
\mathbb{E}_{\mathcal{B}}&\Big[T_{\alpha}\left(P(b,z^n)\parallel p^{U}(b)p(z^n)\right)\Big]\label{Tisalis:(0,1):converse:1}\\
&=\frac{1}{\alpha-1}\,\mathbb{E}_\mathcal{B}\left[\sum_{b,z^n}M^{\alpha-1}p(z^n)P^{\,\alpha}(b\lvert z^n)-1\right]\label{Tisalis:(0,1):converse:2}\\
&=\frac{1}{\alpha-1}\,\mathbb{E}_\mathcal{B}\left[\sum_{z^n}M^{\alpha}p(z^n)P^{\,\alpha}(b=1\lvert z^n)-1\right]\label{Tisalis:(0,1):converse:3}\\
&=\frac{1}{\alpha-1}\,\mathbb{E}_\mathcal{B}\left[\sum_{z^n}p(z^n)\Big(M^{\alpha}P^{\,\alpha}(b=1\lvert z^n)-1\Big)\right]\label{Tisalis:(0,1):converse:4}\\
&=\frac{1}{\alpha-1}\sum_{z^n}p(z^n)\,\mathbb{E}_{\mathcal{B}}\left[M^{\alpha}\left(\sum_{x^n}p(x^n\lvert z^n)\mathds{1}\{\mathcal{B}(x^n)=1\}\right)^{\alpha}-1\right],\label{Tsalis:(0,1):converse:5}
\end{align}

\eqref{Tisalis:(0,1):converse:3} is on account of symmetry, and \eqref{Tsalis:(0,1):converse:5} is the result of substituting \eqref{eqn:rand_bin_dist}. Suppose that $\mathcal{A}_{\epsilon}^n$ is the $\epsilon$-typical set of $z^n$ sequences defined as
\begin{align}
&\mathcal{A}_{\epsilon}^n=\left\{z^n:\Big\lvert\frac{1}{n}\sum_{i=1}^{n}\mathds{1}\{z_i=z\}-p(z)\Big\rvert<\epsilon\right\}.\label{Tisalis:(0,1):converse:6}
\end{align}
Decomposing the sum into typical and non-typical $z^n$ sequences in \eqref{Tsalis:(0,1):converse:5} gives
\begin{align}
\eqref{Tsalis:(0,1):converse:5}&=\frac{1}{\alpha-1}\sum_{z^n\in\mathcal{A}^n_{\epsilon}}p(z^n)\mathbb{E}_{\mathcal{B}}\left[M^{\alpha}\left(\sum_{x^n}p(x^n\lvert z^n)\mathds{1}\{\mathcal{B}(x^n)=1\}\right)^{\alpha}-1\right]\nonumber\\
&\quad+\frac{1}{\alpha-1}\sum_{{z^n\not\in\mathcal{A}^n_{\epsilon}}}p(z^n)\mathbb{E}_{\mathcal{B}}\left[M^{\alpha}\left(\sum_{x^n}p(x^n\lvert z^n)\mathds{1}\{\mathcal{B}(x^n)=1\}\right)^{\alpha}-1\right]\label{Tisalis:(0,1):converse:10}\\
&\geq\frac{1}{\alpha-1}\sum_{z^n\in\mathcal{A}^n_{\epsilon}}p(z^n)\mathbb{E}_{\mathcal{B}}\left[M^{\alpha}\left(\sum_{x^n}p(x^n\lvert z^n)\mathds{1}\{\mathcal{B}(x^n)=1\}\right)^{\alpha}-1\right]\nonumber\\
&\quad+\frac{1}{\alpha-1}\sum_{{z^n\not\in\mathcal{A}^n_{\epsilon}}}p(z^n)\left[\left(M\mathbb{E}_{\mathcal{B}}\sum_{x^n}p(x^n\lvert z^n)\mathds{1}\{\mathcal{B}(x^n)=1\}\right)^{\alpha}-1\right]\label{Tisalis:(0,1):converse:12}\\
&=\frac{1}{\alpha-1}\sum_{z^n\in\mathcal{A}^n_{\epsilon}}p(z^n)\mathbb{E}_{\mathcal{B}}\left[M^{\alpha}\left(\sum_{x^n}p(x^n\lvert z^n)\mathds{1}\{\mathcal{B}(x^n)=1\}\right)^{\alpha}-1\right],\label{eqn:Tsallis_typic_z}
\end{align}
where \eqref{Tisalis:(0,1):converse:12} is a direct consequence of applying Jensen's inequality to the concave function $f(x)=x^{\alpha},\,\alpha\in(0,1)$. Given $z^n$, we define the conditional $\epsilon'$-typical set as
\begin{align}
\nu_n^\ast=\left\{x^n:\Big\lvert\frac{1}{n}\sum_{i=1}^{n}\mathds{1}\{(x_i,z_i)=(x,z)\}-p(x,z)\Big\rvert<\epsilon^\prime\right\}\label{Tisalis:(0,1):converse:23}.
\end{align}
Once again, the approach involves dividing the sum over $x^n$ into two parts: one for typical sequences and the other for non-typical sequences. Building upon \eqref{eqn:Tsallis_typic_z}, we can obtain an upper bound for the expectation term as follows.
\begin{align}
\mathbb{E}_{\mathcal{B}}&\left[M^{\alpha}\left(\sum_{x^n}p(x^n\lvert z^n)\mathds{1}\{\mathcal{B}(x^n)=1\}\right)^{\alpha}-1\right]\label{Tisalis:(0,1):converse:24}\\
&\leq\mathbb{E}_{\mathcal{B}}\left[M^{\alpha}\left(\sum_{x^n\in\nu_n^\ast}p(x^n\lvert z^n)\mathds{1}\{\mathcal{B}(x^n)=1\}\right)^{\alpha}-1\right]\nonumber\\
&\quad+\mathbb{E}_{\mathcal{B}}\left[\left(M\sum_{x^n\not\in\nu_n^\ast}p(x^n\lvert z^n)\mathds{1}\{\mathcal{B}(x^n)=1\}\right)^{\alpha}\right]\label{Tisalis:(0,1):converse:26}\\
&\leq\mathbb{E}_{\mathcal{B}}\left[M^{\alpha}\left(\sum_{x^n\in\nu_n^\ast}p(x^n\lvert z^n)\mathds{1}\{\mathcal{B}(x^n)=1\}\right)^{\alpha}-1\right]\nonumber\\
&\quad+\left(M\mathbb{E}_{\mathcal{B}}\left[\sum_{x^n\not\in\nu_n^\ast}p(x^n\lvert z^n)\mathds{1}\{\mathcal{B}(x^n)=1\}\right]\right)^{\alpha}\label{Tisalis:(0,1):converse:28}\\
&=\mathbb{E}_{\mathcal{B}}\left[M^{\alpha}\left(\sum_{x^n\in\nu_n^\ast}p(x^n\lvert z^n)\mathds{1}\{\mathcal{B}(x^n)=1\}\right)^{\alpha}-1\right]+\left(\sum_{x^n\not\in\nu_n^\ast}p(x^n\lvert z^n)\right)^{\alpha}\label{Tisalis:(0,1):converse:29}\\
&\leq\left(\sum_{x^n\in\nu_n^\ast}\frac{p(x^n\lvert z^n)}{\Big(Mp(x^n\lvert z^n)\Big)^{1-\alpha}}-1\right)+\left(\sum_{x^n\not\in\nu_n^\ast}p(x^n\lvert z^n)\right)^{\alpha},\label{Tisalis:(0,1):converse:31}
\end{align}
where \eqref{Tisalis:(0,1):converse:26} is derived from the inequality
\begin{align}
\left(x+y\right)^{\alpha}\leq x^\alpha+y^\alpha,
\end{align}
which holds for $\alpha\in(0,1)$. On the other hand, \eqref{Tisalis:(0,1):converse:28} follows from Jensen's inequality applied to the concave function $f(x)=x^\alpha$ with $\alpha\in(0,1)$.

By taking into account the sequences $x^n\in\nu_n^\ast$ and conditioning on the knowledge of their binning index, while employing the law of total expectation, it is possible to establish an upper bound for the expectation term in \eqref{Tisalis:(0,1):converse:29}. Consequently, this leads to the derivation of \eqref{Tisalis:(0,1):converse:31} in the following manner.
\begin{align}
&\mathbb{E}_{\mathcal{B}}\left[M^{\alpha}\left(\sum_{x^n\in\nu_n^\ast}p(x^n\lvert z^n)\mathds{1}\{\mathcal{B}(x^n)=1\}\right)^{\alpha}\right]\label{eqn:primary_exp}\\
&=\mathbb{E}_{\mathcal{B}}\left[M^\alpha\left(\sum_{x^n\in\nu_n^\ast}\frac{p(x^n\lvert z^n)\mathds{1}\{\mathcal{B}(x^n)=1\}}{\left(\sum_{\bar{x}^n\in\nu_n^\ast}p(\bar{x}^n\lvert z^n)\mathds{1}\{\mathcal{B}(\bar{x}^n)=1\}\right)^{1-\alpha}}\right)\right]\label{Tisalis:(0,1):converse:30}\\
&=\mathbb{E}_{\mathcal{B}(\nu_n^\ast)}\mathbb{E}_{\mathcal{B}|\mathcal{B}(\nu_n^\ast)}\left[M^\alpha\left(\sum_{x^n\in\nu_n^\ast}\frac{p(x^n\lvert z^n)\mathds{1}\{\mathcal{B}(x^n)=1\}}{\left(\sum_{\bar{x}^n\in\nu_n^\ast}p(\bar{x}^n\lvert z^n)\mathds{1}\{\mathcal{B}(\bar{x}^n)=1\}\right)^{1-\alpha}}\right)\right]\label{Tisalis_totalExp}\\
&=\mathbb{E}_{\mathcal{B}(\nu_n^\ast)}\left[\frac{1}{M}\sum_{x^n\in\nu_n^\ast}\frac{M^\alpha p(x^n\lvert z^n)}{\left(p({x}^n\lvert z^n)+\sum_{x^n\neq\bar{x}^n\in\nu_n^\ast}p(\bar{x}^n\lvert z^n)\mathds{1}\{\mathcal{B}(\bar{x}^n)=1\}\right)^{1-\alpha}}+\frac{M-1}{M}\times 0\right]\label{Tisalis:(0,1):converse:19}\\
&=\mathbb{E}_{\mathcal{B}(\nu_n^\ast)}\left[\sum_{x^n\in\nu_n^\ast}\frac{p(x^n\lvert z^n)}{M^{1-\alpha}\left(p(x^n\lvert z^n)+\sum_{x^n\neq \bar{x}^n\in\nu_n^\ast}p(\bar{x}^n\lvert z^n)\mathds{1}\{\mathcal{B}(\bar{x}^n)=1\}\right)^{1-\alpha}}\right]\label{Tisalis:(0,1):converse:20}\\
&\leq\sum_{x^n\in\nu_n^\ast}\frac{p(x^n\lvert z^n)}{\Big(Mp(x^n\lvert z^n)\Big)^{1-\alpha}}.\label{Tisalis:(0,1):converse:32}
\end{align}
\eqref{Tisalis:(0,1):converse:30} is simply a restatement of \eqref{eqn:primary_exp}. In \eqref{Tisalis_totalExp}, we utilize the law of total expectation by taking into account that expectations are applied to each $x^n\in\nu_n^\ast$ individually. Then, by applying $\mathbb{E}_{\mathcal{B}|\mathcal{B}(x^n)}$ step-by-step to each sequence $x^n\in\nu_n^\ast$ and considering that $\mathcal{B}$ is a randomly uniformly mapping, we arrive at \eqref{Tisalis:(0,1):converse:19}. Further simplification and the elimination of the summation term in the denominator yield equation \eqref{Tisalis:(0,1):converse:32}.

By combining equations \eqref{eqn:Tsallis_typic_z} and \eqref{Tisalis:(0,1):converse:31}, and considering that $\alpha$ is within the range of $(0,1)$, we obtain the following lower bound for Tsallis divergence.
\begin{align}
\mathbb{E}_{\mathcal{B}}\Big[T_{\alpha}\left(P(b,z^n)\parallel p^{U}(b)p(z^n)\right)\Big]&\geq\frac{1}{\alpha-1}\sum_{z^n\in\mathcal{A}^n_{\epsilon}}p(z^n)\left(\sum_{x^n\in\nu_n^\ast}\frac{p(x^n\lvert z^n)}{\Big(Mp(x^n\lvert z^n)\Big)^{1-\alpha}}-1\right) \label{eqn:LB_Tsallis_less1}\\
&\qquad+\frac{1}{\alpha-1}\sum_{{z^n\in\mathcal{A}^n_{\epsilon}}}p(z^n)\left(\sum_{x^n\not\in\nu_n^\ast}p(x^n\lvert z^n)\right)^{\alpha}\label{Tisalis:(0,1):converse:33}.
\end{align}
For every sequence $x^n \in \nu_n^\ast$, we have the inequality $p(x^n\lvert z^n) \geq 2^{-n(H(X\lvert Z)+\epsilon^{\prime\prime})}$. Hence, when we set $M=2^{nR}$ and choose $R>H(X\lvert Z)+\epsilon^{\prime\prime}$, the right side of \eqref{eqn:LB_Tsallis_less1} converges to $1/(1-\alpha)$, which is strictly positive. This establishes the completion of the proof.
\end{proof}
\subsection{An asymptotic analysis of the OSRB rate for a special type}
\label{appendix:lemma_D}	
	
\begin{lemma}
\label{asymptotic:analysis:on:rate:type}
Let $X^n$ be distributed over $\mathcal{T}_{\epsilon}^n(p_X)$ with the distribution
\begin{align}
\tilde{p}(x^n)=\frac{\prod_{i=1}^{n}p(x_i)}{\sum_{x^n\in\mathcal{T}_{\epsilon}^n(p_X)}\prod{_{i=1}^{n}}p(x_i)},
\end{align}
and $p_{Z^n\lvert X^n}=\prod_{i=1}^{n}p_{Z_i\lvert X_i}$. Then, for any $\alpha\in(1,\infty)$ and $k\in\{0,\cdots,\alpha-1\}$, we have
\begin{align}
\lim_{n\to\infty}\frac{1}{n(\alpha-k-1)}&\log\left(\sum_{z^n,\,x^n\in\mathcal{T}_{\epsilon}^{n}(p_X)}q(z^n)\frac{p^{\alpha-k}(x^n,z^n)}{q^{\alpha-k}(z^n)}\right)\nonumber\\
&\leq -H(X)+\sum_{x}p(x)\,D_{\alpha-k}\Big(p(z\lvert x)\parallel p(z)\Big),\label{Asymp:OSRB:analys:rate:type:2}
\end{align}
where $q(z^n)=\prod_{i=1}^{n}p(z_i)$.

\end{lemma}

\begin{proof}
\begin{align}
&\log\left(\sum_{z^n,\,x^n\in\mathcal{T}_{\epsilon}^n(p_X)}q(z^n)\frac{p^{\alpha-k}(x^n,z^n)}{q^{\alpha-k}(z^n)}\right)\label{Asymp:OSRB:analys:rate:type:3}\\
&\quad\leq-n(\alpha-k-1)\left(H(X)-\delta_n(\epsilon)\right)+\log\left(\sum_{z^n,x^n\in\mathcal{T}_{\epsilon}^n(p_X)}q(z^n)\tilde{p}(x^n)\frac{p^{\alpha-k}(z^n\lvert x^n)}{q^{\alpha-k}(z^n)}\right)\label{Asymp:OSRB:analys:rate:type:4}\\
&\quad\leq-n(\alpha-k-1)\left(H(X)-\delta_n(\epsilon)\right)+\log\left(\sum_{z^n}q(z^n)\frac{p^{\alpha-k}(z^n\lvert \bar{x}^n)}{q^{\alpha-k}(z^n)}\right)\label{Asymp:OSRB:analys:rate:type:5}\\
&\quad=-n(\alpha-k-1)\left(H(X)-\delta_n(\epsilon)\right)+\log\left(\sum_{z^n}\prod_{i=1}^{n}p(z_i)\frac{p^{\alpha-k}(z_i\lvert \bar{x}_i)}{p^{\alpha-k}(z_i)}\right)\label{Asymp:OSRB:analys:rate:type:6}\\
&\quad\leq-n(\alpha-k-1)\left(H(X)-\delta_n(\epsilon)\right)+\log\prod_{x}\left(\sum_{z}p(z)\frac{p^{\alpha-k}(z\lvert{x})}{p^{\alpha-k}(z)}\right)^{n(p(x)+\epsilon)}\label{Asymp:OSRB:analys:rate:type:7}\\
&\quad=-n(\alpha-k-1)\left(H(X)-\delta_n(\epsilon)\right)+n\sum_{x}(p(x)+\epsilon)\log\left(\sum_{z}p(z)\frac{p^{\alpha-k}(z\lvert{x})}{p^{\alpha-k}(z)}\right)\label{Asymp:OSRB:analys:rate:type:8}\\
&\quad=-n(\alpha-k-1)\left(H(X)-\delta_n(\epsilon)\right)+n(\alpha-k-1)\sum_{x}(p(x)+\epsilon)D_{\alpha-k}\Big(p(z\lvert x)\parallel p(z)\Big).\label{Asymp:OSRB:analys:rate:type:9}
\end{align}
\eqref{Asymp:OSRB:analys:rate:type:4} utilizes $\frac{1}{n}\log(\tilde{p}(x^n))\leq-H(X)+\delta_n(\epsilon)$ for $x^n\in\mathcal{T}_{\epsilon}^n(p_X)$. In \eqref{Asymp:OSRB:analys:rate:type:5}, we fix $x^n=\bar{x}^n$ as follows:

\begin{align}
\sum_{z^n,\,x^n\in\mathcal{T}_{\epsilon}^n(p_X)}q(z^n)\tilde{p}(x^n)\frac{p^{\alpha-k}(z^n\lvert x^n)}{q^{\alpha-k}(z^n)}&\leq\max_{x^n\in\mathcal{T}_{\epsilon}^n(p_X)}\sum_{z^n}q(z^n)\frac{p^{\alpha-k}(z^n\lvert x^n)}{q^{\alpha-k}(z^n)}\label{simpification:binning:rate:type:1}\\
&=\sum_{z^n}q(z^n)\frac{p^{\alpha-k}(z^n\lvert \bar{x}^n)}{q^{\alpha-k}(z^n).}\label{simpification:binning:rate:type:2}
\end{align}
Finally, letting $n\to\infty$ and $\epsilon\to 0$, and thus $\delta_n(\epsilon),\delta_n\to 0$, completes the proof.

\end{proof}


\subsection{\texorpdfstring{$T_\alpha$ and $D_{\infty}$}{} on conditional probability distributions}
\label{subsec:useful_lemma}

In the following Lemma, we show how $T_\alpha$ and $D_{\infty}$ on the conditional distribution is related to $T_\alpha$ on the joint distribution.

\begin{lemma}
\label{lemma:D_cond}
Consider an arbitrary distribution $q(z^n)$, let $\alpha \in (1, \infty)$, and assume that
\begin{align}
\label{eqn:D_alpha_joint}
\mathbb{E}_\mathcal{B}\left[T_\alpha\Big(P(m,f,z^n)\parallel p^{U}\!(m)\,p^{U}\!(f)\,q(z^n)\Big)\right]\leq\epsilon,
\end{align}
then with probability at least $\frac{1}{2}$ over all pair $(\mathcal{B},F)$,
\begin{align}
	\label{eqn:D_alpha_joint_2}
T_\alpha\Big(P(m,z^n|f)\parallel p^{U}\!(m)\,q(z^n)\Big)\leq \eta(\epsilon),
\end{align}
where $\eta(\epsilon)$ is a constant depending only on $\epsilon$ and it vanishes as $\epsilon$ vanishes.
\end{lemma}
\begin{proof}
According to the contractivity property of $T_\alpha$ for $\alpha\geq 1$, the following result is obtained from \eqref{eqn:D_alpha_joint}.
\begin{align}
\mathbb{E}_\mathcal{B}\left[T_{\alpha}\Big(P(f)\parallel p^{U}\!(f)\Big)\right]\leq\epsilon.
\end{align}
Then the non-decreasing property of $T_\alpha$ with $\alpha$ results
\begin{align}
\mathbb{E}_\mathcal{B}\left[D\Big(P(f)\parallel p^{U}\!(f)\Big)\right]\leq\epsilon.
\end{align}
In addition, by applying Pinsker's inequality and Jensen's inequality, we have
\begin{align}
\label{eqn:TV_pinsker}
\mathbb{E}_{\mathcal{B}}\Big[\|P(f)-p^{U}\!(f)\|_{\mathrm{TV}}\Big]\leq c\sqrt{\epsilon},
\end{align} 
where $c$ is a constant and it is equal to $\sqrt{\frac{\ln 2}{2}}$ if we take two as the base of logarithm. From here, with a probability of at least $\frac{7}{8}$, we have
\begin{align}
\label{eqn:TV_pinsker-y}
\|P(f)-p^{U}\!(f)\|_{\mathrm{TV}}\leq 8c\sqrt{\epsilon},
\end{align} 
Further,   the assumption \eqref{eqn:D_alpha_joint} implies that with probability at least $\frac{7}{8}$, we have
\begin{equation}
T_\alpha\Big(p(m,f,z^n)\parallel p^{U}\!(m)\,p^{U}\!(f)\,q(z^n)\Big)\leq 8\epsilon.\label{eq:yas-20}
\end{equation}
Thus with with probability at least $\frac{3}{4}$ both the inequalities \eqref{eqn:TV_pinsker-y} and \eqref{eq:yas-20} are simultaneously satisfied. Let $\beta$ is one of those codebooks for which both the inequalities \eqref{eqn:TV_pinsker-y} and \eqref{eq:yas-20} are simultaneously satisfied.

Representing the total variation distance as a $f$-divergence, we can rewrite \eqref{eqn:TV_pinsker-y} as below,
\begin{align}
\mathbb{E}_{F}\left[\Big\lvert\frac{p^{U}\!(F)}{P (F)}-1\Big\rvert\right]\leq 8c\sqrt{\epsilon},
\end{align}
which with a probability of at least $1-\sqrt{8c\sqrt{\epsilon}}$ (over random variable $F$ conditioned on fixed binning $\beta$) gives
\begin{align}\label{T_alpha_proprtion_relizations}
\Big\lvert\frac{p^{U}\!(f)}{p(f)}-1\Big\rvert\leq \sqrt{8c\sqrt{\epsilon}}.
\end{align}
In other words, by defining $\delta(\epsilon):= \sqrt{8c\sqrt{\epsilon}}$, we have with probability $\ge 1-\delta(\epsilon)$, 
\begin{align}
\label{eqn:bound_f}
1-\delta(\epsilon)\leq\frac{p^{U}\!(f)}{p(f)}\leq 1+\delta(\epsilon).
\end{align}
Let $\mathcal{G}$ be the set of all $f$ for which \eqref{eqn:bound_f} holds.
Further \eqref{eq:yas-20} implies
\begin{align}
\label{eqn:ineq_uncond}
1+8(\alpha-1)\epsilon&\geq 
\sum_{m,z^n,f}\frac{p^{\alpha}(m,z^n,f)}{\left(\frac{1}{M}p^{U}\!(f)q(z^n)\right)^{\alpha-1}}.
\end{align}
Using the definition of $\mathcal{G}$, we have
\begin{align}
1+8(\alpha-1)\epsilon&\geq 
\sum_{m,z^n,f}\frac{p^{\alpha}(m,z^n,f)}{\left(\frac{1}{M}p^{U}\!(f)q(z^n)\right)^{\alpha-1}}\\
& \ge \sum_{(m,z^n,f):f\in\mathcal{G}}\frac{p^{\alpha}(m,z^n,f)}{\left(\frac{1}{M}p^{U}\!(f)q(z^n)\right)^{\alpha-1}}\\
&=\sum_{(m,z^n,f):f\in\mathcal{G}}p(f)\frac{p^{\alpha}(m,z^n|f)}{\left(\frac{1}{M}q(z^n)\right)^{\alpha-1}}\cdot\left(\frac{p(f)}{p^{U}\!(f)}\right)^{\alpha-1}\\
&\ge\frac{1}{(1+\delta(\epsilon))^{\alpha-1}}\sum_{(m,z^n,f):f\in\mathcal{G}}p(f)\frac{p^{\alpha}(m,z^n|f)}{\left(\frac{1}{M}q(z^n)\right)^{\alpha-1}}\\
&=\frac{1}{(1+\delta(\epsilon))^{\alpha-1}}\mathbb{E}_F\left[\sum_{(m,z^n)}\frac{p^{\alpha}(m,z^n|F)}{\left(\frac{1}{M}q(z^n)\right)^{\alpha-1}}\mathds{1}\{F\in\mathcal{G}\}\right]\\
&=\frac{1}{(1+\delta(\epsilon))^{\alpha-1}}\mathbb{E}_F\left[\left(1+(\alpha-1)    T_\alpha\Big(p(m,z^n|F)\parallel p^{U}\!(m)\,q(z^n)\Big)\right)\mathds{1}\{F\in\mathcal{G}\}\right]\label{eqn:y-t-f}
\end{align}
Define $\tilde{\delta}(\epsilon):=(1+\delta(\epsilon))^{\alpha-1}(1+8(\alpha-1)\epsilon)-1$. Observe that 
$\tilde{\delta}(\epsilon)$ vanishes as $\epsilon$ vanishes. Rearranging \eqref{eqn:y-t-f} implies,
\begin{align}
(\alpha-1) \mathbb{E}_F\left[\left(   T_\alpha\Big(p(m,z^n|F)\parallel p^{U}\!(m)\,q(z^n)\Big)\right)\mathds{1}\{F\in\mathcal{G}\}\right]\le 1+\tilde{\delta}(\epsilon)-\mathbb{P}[F\in\mathcal{G}]\le \delta(\epsilon)+\tilde{\delta}(\epsilon)
\end{align}
Thus with probability at least $\frac{3}{4}-\mathbb{P}[F\notin\mathcal{G}]\ge \frac{2}{3}$ (for small enough $\epsilon$), we have
\begin{align}
T_\alpha\Big(p(m,z^n|F)\parallel p^{U}\!(m)\,q(z^n)\Big)\le \eta(\epsilon)\label{eqn:occ0}
\end{align}
where $\eta(\epsilon):=\frac{4}{(\alpha-1)}( \delta(\epsilon)+\tilde{\delta}(\epsilon))$. Finally the probability of occurring \eqref{eqn:occ0} over the pair $(\mathcal{B},F)$ is at least the probability that \eqref{eqn:TV_pinsker-y}, \eqref{eq:yas-20}  and \eqref{eqn:occ0} occurring simultaneously. By Bayes rule, this probability is at least $\frac{3}{4}\cdot\frac{2}{3}=\frac{1}{2}$.

This statement concludes the proof.
\end{proof}
	\begin{remark}
		We can obtain a similar result to Lemma \ref{lemma:D_cond} for $D_{\infty}$ instead of $T_{\alpha}$ by repeating the steps with the parameters
\begin{align}
    \eta^{\prime}(\epsilon) = 8\epsilon + \log\left(1 + \delta(\epsilon)\right),
\end{align}
where $\eta^{\prime}(\epsilon)$ replaces $\eta(\epsilon)$, and $\delta(\epsilon) = \sqrt{8c\sqrt{\epsilon}}$ is defined in Lemma \ref{lemma:D_cond}.
	\end{remark}

\begin{lemma}
\label{lemma:tsallis_mess}
Let us consider an arbitrary distribution of $q(z^n)$. Suppose there are two distributions over the variables $(m,f,x^n,z^n)$ given by
\begin{align*}
\circled{\rm{1}}&\qquad p(m|f)\,p(x^n|m,f)\,p(z^n|x^n),\\
\circled{\rm{2}}&\qquad p^U(m)\,p(x^n|m,f)\,p(z^n|x^n).
\end{align*}
Assume that the following condition holds for distribution $\circled{\rm{1}}$.
\begin{align*}
T_\alpha\Big(p_1(m,z^n|f)\parallel p^{U}\!(m)\,q(z^n)\Big)\le \xi.
\end{align*}
Then, there exists $M'\in\mathcal{M}'$ with $|\mathcal{M}'|\geq|\mathcal{M}|/2$, such that for distribution in $\circled{\rm{2}}$,
\begin{align*}
T_\alpha\Big(p_2(m',z^n|f)\parallel p^{U}\!(m')\,q(z^n)\Big)\le \delta(\xi).
\end{align*}
Moreover, the function $\delta(\xi)$ converges to zero as $\xi$ approaches zero.
\end{lemma}

\begin{proof}
Given that
\begin{align}
\label{eqn:tsallis_p1}
T_\alpha\Big(p_1(m,z^n|f)\parallel p^{U}\!(m)\,q(z^n)\Big)=\frac{1}{\alpha-1}\mathbb{E}_{M,Z^n}\Bigg[\left(\frac{p_1(M,Z^n\lvert F=f)}{p^{U}(M)\,q(Z^n)}\right)^{\alpha}-1\Bigg]\le \xi,
\end{align} 
we can deduce, by employing Markov's inequality, that a minimum of $3/4$ of the instances $m\in\mathcal{M}$, satisfy
\begin{align}
\label{simplification:T_alpha_condition:2}
\mathbb{E}_{Z^n}\Bigg[\left(\frac{p_1(m,Z^n\lvert F=f)}{p^{U}(m)\,q(Z^n)}\right)^\alpha\Bigg]\leq 1+4(\alpha-1)\xi.
\end{align}

Furthermore, it can be deduced from \eqref{eqn:tsallis_p1}, by combining the contractivity property of the Tsallis divergence with Pinsker's inequality, that
\begin{align}
\label{simplification:ratio:3/4:f:2}
\mathbb{E}_{M}\Bigg[\bigg\lvert\frac{p(M\lvert F=f)}{p^{U}(M)}-1\bigg\rvert\Bigg]=\Big\|p(m\lvert F=f)-p^{U}\!(m)\Big\|_{\mathrm{TV}}\leq c\sqrt{\xi},
\end{align}
where $c=\sqrt{\ln 2/2}$. Thus, employing Markov's inequality once again, it can be inferred that for over $3/4$ of realizations of $m\in\mathcal{M}$, the following inequality holds.
\begin{align}
\label{simplification:ratio:3/4:f:1}
1-4c\sqrt{\xi}\leq\frac{p(m\lvert F=f)}{p^{U}(m)}\leq 1+4c\sqrt{\xi}
\end{align}
Hence, for a minimum of half of the $\mathcal{M}$ members, both constraints \eqref{simplification:T_alpha_condition:2} and \eqref{simplification:ratio:3/4:f:1} are met. By selecting these qualified members and assigning them to the set $\mathcal{M}'$ with $|\mathcal{M}'|\geq|\mathcal{M}|/2$, we obtain
\begin{align}
&T_\alpha\Big(p_2(m',z^n|f)\parallel p^{U}\!(m')\,q(z^n)\Big)\nonumber\\
&\qquad=\frac{1}{\alpha-1}\mathbb{E}_{M^\prime, Z^n}\Bigg[\left(\frac{p_2(M^\prime,Z^n\lvert F=f)}{p_1(M^\prime,Z^n\lvert F=f)}\right)^{\alpha}\left(\frac{p_1(M^\prime,Z^n\lvert F=f)}{p^{U}(M^\prime)\,q(Z^n)}\right)^{\alpha}-1\Bigg]\label{simplification:T_alpha_condition:6}\\
&\qquad=\frac{1}{\alpha-1}\left[\mathbb{E}_{M^\prime, Z^n}\Bigg[\left(\frac{p^{U}(M^\prime)}{p(M^\prime\lvert F=f)}\right)^{\alpha}\left(\frac{p_1(M^\prime,Z^n\lvert F=f)}{p^{U}(M^\prime)\,q(Z^n)}\right)^{\alpha}\Bigg]-1\right]\\
&\qquad\leq\frac{1}{\alpha-1} \left[\left({1+4(\alpha-1)\xi}\right)\left(\frac{1}{1-4c\sqrt{\xi}}\right)^\alpha-1\right]=\delta(\xi)\label{eqn:cons2_ts},
\end{align}
where \eqref{eqn:cons2_ts} are derived from \eqref{simplification:ratio:3/4:f:1} and \eqref{simplification:T_alpha_condition:2}, respectively. 

\end{proof}

 \begin{remark}
Similar to Lemma \ref{lemma:tsallis_mess}, this holds for $D_{\infty}$ instead of $T_{\alpha}$ with 
\begin{align}
    \delta^\prime(\xi) = \xi + \log\left(\frac{1}{1 - 2c\sqrt{\xi}}\right),
\end{align} 
where $\delta(\xi)$ is replaced with $\delta^\prime(\xi)$, and $c = \sqrt{\frac{\ln 2}{2}}$.
 \end{remark}

\begin{remark}\label{stochstic:uniform:messages}
	For the generalization of Lemma \ref{lemma:tsallis_mess} from a deterministic encoder to a stochastic encoder, we consider new distributions 
\begin{align}
    \circled{\rm{1}}&\qquad p(m|f)\,\tilde{p}(u^n|m,f)\,\tilde{p}(x^n|u^n)\,p(z^n|x^n),\\
    \circled{\rm{2}}&\qquad p^U(m)\,\tilde{p}(u^n|m,f)\,\tilde{p}(x^n|u^n)\,p(z^n|x^n),
\end{align}
over the variables $(m,f,u^n,x^n,z^n)$. The proof has similar steps to that of Lemma \ref{lemma:tsallis_mess}.
\end{remark}

\end{document}